\documentclass{article}

\usepackage{arxiv}

\usepackage[utf8]{inputenc} % allow utf-8 input
\usepackage[T1]{fontenc}    % use 8-bit T1 fonts
\usepackage{hyperref}       % hyperlinks
\usepackage{url}            % simple URL typesetting
\usepackage{booktabs}       % professional-quality tables
\usepackage{amsfonts}       % blackboard math symbols
\usepackage{nicefrac}       % compact symbols for 1/2, etc.
\usepackage{microtype}      % microtypography
\usepackage{graphicx}
\usepackage{natbib}
\usepackage{doi}
\usepackage{multicol}

\title{3D Anatomical Representations and Analysis: an Application to the Spine}

\date{} 					% Or removing it

\author{{\hspace{1mm}Martina~Paccini} \\
	Istituto di Matematica Applicata \\e Tecnologie Informatiche\\ `E. Magenes' CNR\\
	Via de Marini 6 \\ Genova 16149 , Italy\\
	\texttt{martina.paccini@ge.imati.cnr.it} \\
	\And
	{\hspace{1mm}Giuseppe~{Patan{\'e}}} \\
    Istituto di Matematica Applicata \\e Tecnologie Informatiche \\`E. Magenes' CNR\\
	Via de Marini 6 \\ Genova 16149 , Italy\\
	\And
	{\hspace{1mm}Michela~Spagnuolo} \\
    Istituto di Matematica Applicata \\e Tecnologie Informatiche \\`E. Magenes' CNR\\
	Via de Marini 6 \\ Genova 16149 , Italy\\
}

\renewcommand{\headeright}{Technical Report}
\begin{document}
\maketitle

\begin{abstract}
This work proposes a framework for the patient-specific characterization of the spine, which integrates information on the tissues with geometric information on the spine morphology. Key elements are the extraction of 3D patient-specific models of each vertebra and the intervertebral space from 3D CT images, the segmentation of each vertebra in its three functional regions, and the analysis of the tissue condition in the functional regions based on geometrical parameters.  The localization of anomalies obtained in the results and the proposed visualization support the applicability of our tool for quantitative and visual evaluation of possible damages, for surgery planning, and early diagnosis or follow-up studies. Finally, we discuss the main properties of the proposed framework in terms of characterisation of the morphology and pathology of the spine on benchmarks of the spine district.
\end{abstract}

% keywords can be removed

\section{Introduction}
The spine is a fundamental element of human anatomy that provides support for our body and organs and guarantees a wide range of mobility. Moreover, the spine deals with load transfer and protects the spinal cord from injuries. All these different tasks can be accomplished thanks to the complex anatomical structure of each element composing the spine and to the organization of the different tissues in the overall district. Biomechanical alterations can lead to pain and disability or even worse consequences. For these reasons, the study of the spine should not be limited to considering either the tissue conditions or the morphology of vertebral elements.

\paragraph{Goals}

 The aim of our work is the 3D characterization of the human spine from a patient-specific and subject-specific perspective, to support the diagnosis of possible pathologies and to improve the visualization of the district when planning surgical interventions. Our framework (Sect.~\ref{Method}) is composed of 4 sub-parts: (i) \emph{extraction of patient-specific 3D models of the whole spine} by combining 3D images and 3D geometry analysis. This evaluation can be extended to a follow-up analysis to compare the subject's evolution in time or different acquisitions (Sect.~\ref{Sect:3DModel}). Then, we address the (ii) \emph{segmentation of the 3D model in the three main functional parts} that constitute the vertebral bones, performing a shape segmentation on the 3D patient-specific model and including a characterization of the tissue status in the neighbourhood of the vertebral surface (Sect.~\ref{Subsect:spineSegm}). This segmentation allows the (iii) \emph{extraction of the intervertebral space} leveraging the information retrievable from the previous analysis (Sect.~\ref{sect:IntervertebralExtraction}). Finally, we discuss the (iv) \emph{quantitative evaluation of the tissue inside the vertebral body} through the computation of HU parameters in an automatic and patient-specific way (Sect.~\ref{Sect:Tissue&Bone}). 

A key element of the framework is an effective and interactive representation of the anatomical structures to facilitate the interpretation of the morphology and pathology of the input district (Sect.~\ref{sec:RESULTS}). A 3D visualization supports the analysis and rendering of morphological changes, but it cannot provide information on the tissue composition or variation over time. In contrast, focusing only on the image intensities does not provide morphological information as detailed as the 3D model's geometrical analysis. To overcome these limitations, we implement a combination of image texture and shape analysis, where the visualization of the 3D image is augmented with the help of the 3D patient-specific models extracted from the image segmentation. Superimposing 3D segmented models of the spine with the underlying volumetric images allows the visualization of the bony/soft tissue structures. In this way, the physicians/surgeons can easily localize the region of interest for further analysis. Moreover, an enhanced visualization is achieved by mapping the HU values of the original image volume on the segmented district~\citep{paccini2020analysis}. Both visualization options combine information extracted from the 3D images and the 2D and 3D information of the patient-specific models. Finally, we discuss the experimental results of the proposed approach on benchmarks of the spine district~\citep{SEKUBOYINA2021102166}, thus showing the main properties of the proposed framework in terms of characterisation of the morphology and pathologies of the spine (Sect.~\ref{sec:DISCUSSION}).

\paragraph{Contribution and novelties} 
As main contribution and novelties, the proposed characterization is patient-specific and supports the early diagnosis of possible pathology and an accurate evaluation of the patient status. Instead of focusing on a particular segment of the spine, our approach is general enough to consider the whole spine. The 3D images are visualized together with 3D patient-specific models, thus helping the navigation of the 3D CT volume.

Our work analyzes both the morphology, through the study of the 3D models, and the tissue conditions of the vertebral spine, based on the HU values of the CT images. In this way, the analysis and characterization of the single subject include both aspects, in contrast to previous work that focuses on either morphology or tissues. Furthermore, the tissue evaluation is performed around the vertebral surface and inside the vertebral body. The latter introduces an automatic selection of a 3D ROI inside each vertebral body, differently from what is usually performed in practice, where a 2D ROI is manually identified by experts. Finally, our characterization includes a combination of geometrical and tissue-related parameters in different analyses. 
\begin{figure}[t]
\centering
(a)\includegraphics[width=0.45\textwidth]{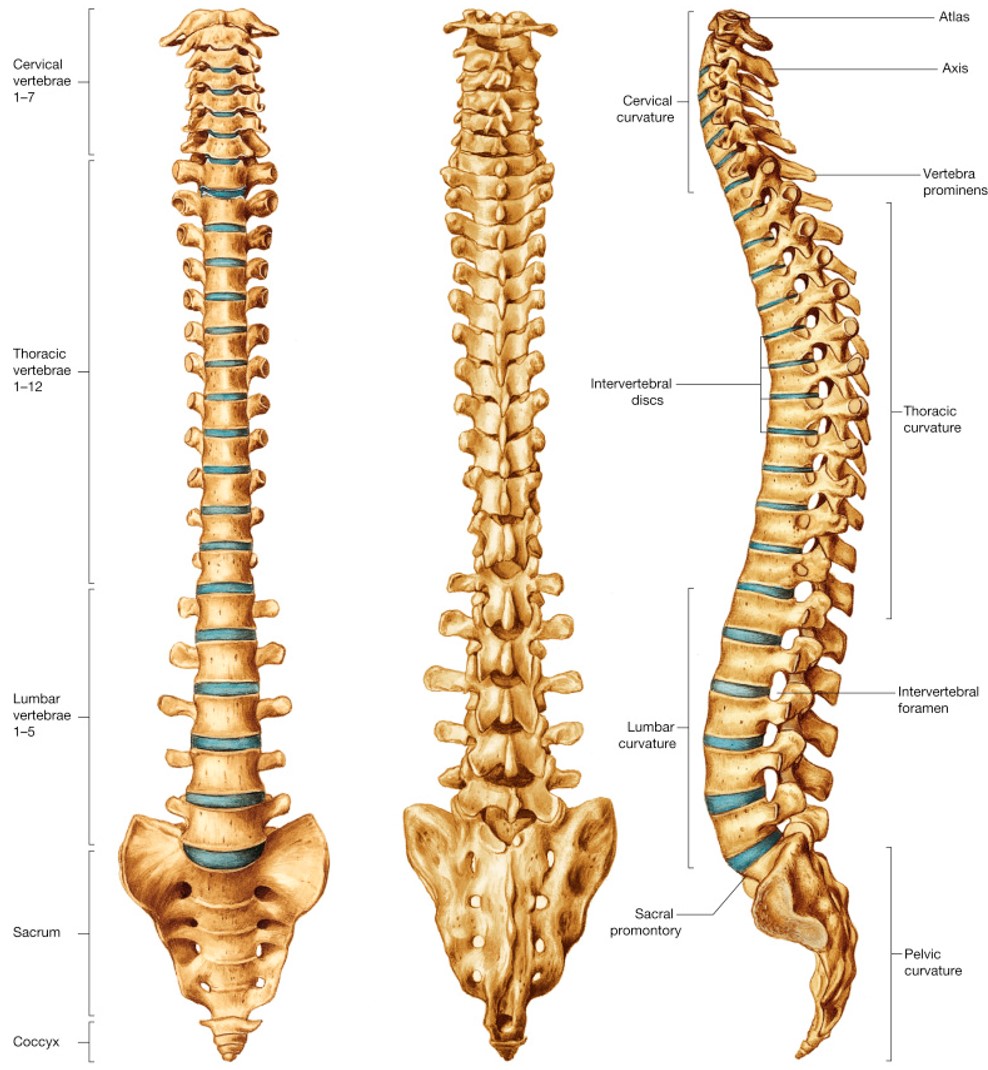}\\
\begin{tabular}{cc}
\includegraphics[height=0.15\textwidth]{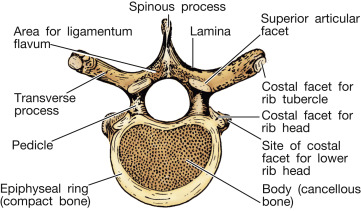}
&\includegraphics[height=0.15\textwidth]{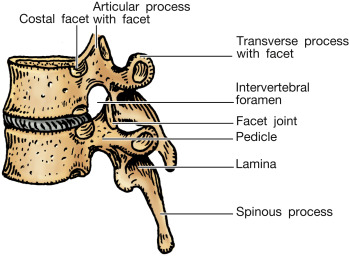}\\
\multicolumn{2}{c}{(b)}
\end{tabular}
\caption{\label{fig:SpineAnatomy} Anatomical description of (a) the spine segments and physiological curves and (b) the vertebral bone and of a vertebral segment (two consecutive vertebrae with the relative intervertebral space. Image courtesy of~\citep{MAHADEVAN2018327}.}
\end{figure}
\section{Previous work\label{sec:PREVIOUS-WORK}}
We review previous work on the morphological characterisation of the spine and its underlying pathologies (Sect.~\ref{sec:MORPHO-PATHO-SPINE}), based on imaging and 3D modelling techniques (Sect.~\ref{sec:IMAGE-MODEL-SPINE}).

\subsection{Morphology and pathologies of the spine\label{sec:MORPHO-PATHO-SPINE}}
\paragraph{Morphology}
As a core component of the muscle-skeletal system, the spine sustains and supports the body and organs, and plays a significant role in mobility and load transfer. The vertebral spine is composed of 33 individual bones that interlock with each other to form the spinal column. Thanks to its structure, the spine also shields the spinal cord from injuries and possible mechanical shocks that arise from impacts~\citep{SEKUBOYINA2021102166}. Indeed, the vertebral column presents the \textit{vertebral canal}, which contains the spinal cord and works as a protection from external trauma. To provide support, powerful muscles insert in the posterior portion of the vertebral column and allow the maintenance of the posture of the subject~\citep{MAHADEVAN2018327}.

For descriptive purposes, the vertebral column is divided into five regions. From above downwards these are, in sequence, the cervical, thoracic, lumbar, sacral, and coccygeal regions. However, only the top 24 bones are movable, meaning that the individual vertebrae can move locally, while the vertebrae of the sacrum and coccyx are fused and, therefore, remain fixed. To perform their main function, the vertebrae present unique features in each region. When viewed from the side, an adult spine has a natural S-shaped curve. The neck (\textit{cervical}) and low back (\textit{lumbar}) regions have a slight concave curve, while the thoracic and sacral regions have a gentle convex curve. The curves absorb shock like a coiled spring, maintain balance, and guarantee the range of motion throughout the spinal column~\citep{MAHADEVAN2018327,SpineAna11:online}. Fig.~\ref{fig:SpineAnatomy}(a) shows the different segments and components of adults' spinal cords.
 \begin{figure}
\centering
\begin{tabular}{ccc}
     \includegraphics[width=0.14\textwidth]{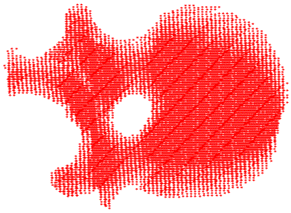} 
     &\includegraphics[width=0.14\textwidth]{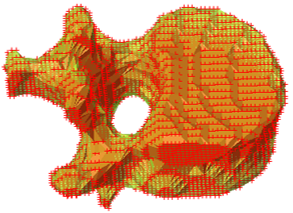}
     &\includegraphics[width=0.14\textwidth]{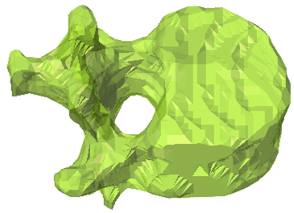}\\
     (a) &(b) &(c)
\end{tabular}
\caption{\label{fig:alhashape}3D vertebral surfaces: (a) input points; (b) vertebral~$\alpha$-shape extracted with the points in (a), and (c) final 3D vertebral surface.}
\end{figure}
\paragraph*{Vertebrae}
The vertebrae have a shape that allows them to interlock with each other to compose the spine. Except for the first two cervical vertebrae, all the other movable vertebrae have similar morphology, even if they present unique features in each region of the spine to perform their main functions. Each vertebra has three functional parts: the \textit{vertebral body} appointed to load-bearing, the \textit{vertebral arch} that protects the spinal cord and \textit{transverse processes} for ligament attachment. The vertebral body is a cylindroid placed anteriorly. The vertebral arch is a bony arch attached to the back of the vertebral body. The transverse processes project laterally from the vertebral arch on both sides, while the spinous process projects backwards from the posterior midline of the vertebral arch. Between the vertebral body and the vertebral arch is the vertebral foramen. Considering the whole spine, all the vertebral foramina are stacked one on the other to build the vertebral canal. The spinal cord occupies the upper two-thirds of the vertebral canal (Fig.~\ref{fig:SpineAnatomy}(a)). The pedicle is the part of the vertebral arch that intersects the back of the vertebral body. Since the pedicle's height is approximately half the height of the vertebral body, there is a gap between the two successive pedicles. This gap is called \emph{intervertebral foramen} and its role is to transmit the spinal nerves with articular arteries and veins~\citep{MAHADEVAN2018327}.
 \begin{figure}[t]
\centering
\centering
\begin{tabular}{cc}
\includegraphics[width=0.4\textwidth]{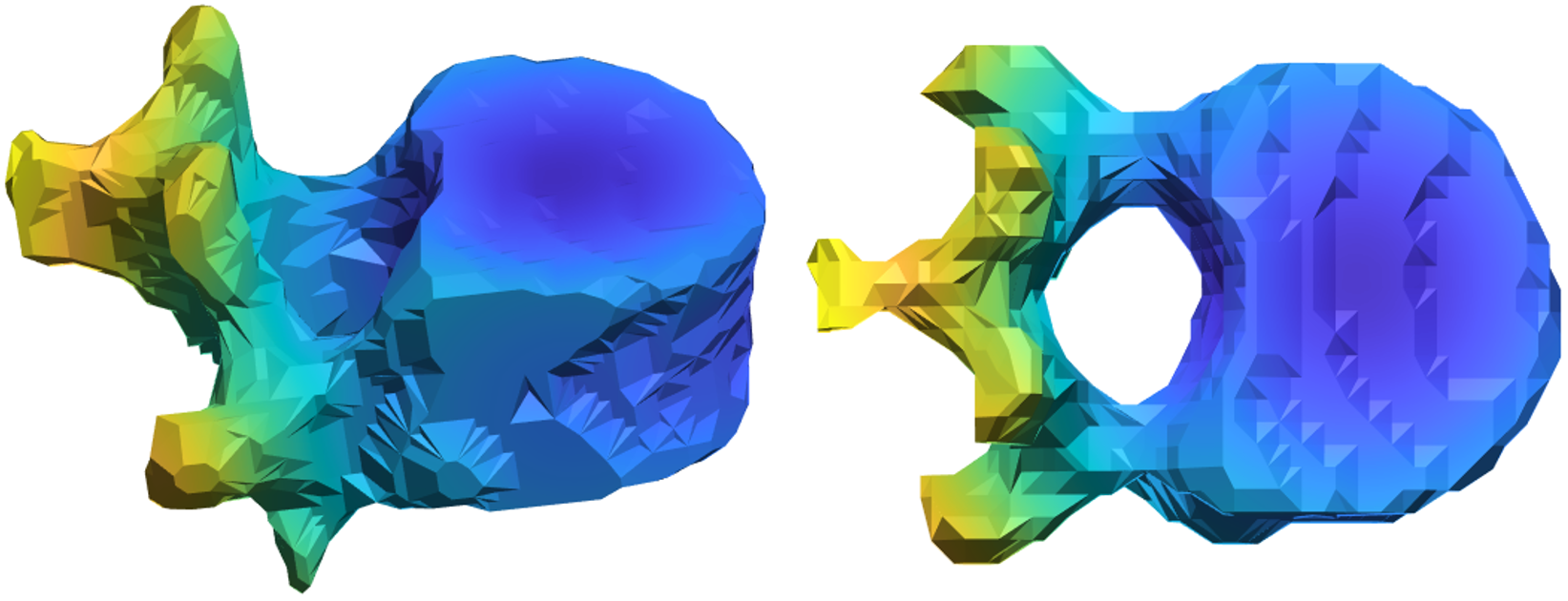}&
\includegraphics[height=0.20 \linewidth]{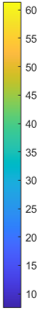}
\end{tabular}
\caption{\label{fig:distanceDistribution}Distance distribution of a vertebral bone. Each vertex is associated with the value of its distance from the vertebral body centroid. The unit measure of the colormap is millimeters.}
\end{figure}
\paragraph{Pathologies}
Biomechanical alterations, in the short term, can lead to pain and disability and, in the long term, even to worse consequences~\citep{SEKUBOYINA2021102166}. Moreover, a damage to the spinal cord can result in a loss of sensory and motor function below the level where the damage presented, e.g., an injury to the cervical area may cause sensory and motor impairment of the arms and legs (\textit{tetraplegia})~\citep{SpineAna11:online}. Given the great relevance and influence of the spine, various efforts have been made to quantify and understand its biomechanics leading to various research branches that focus on the vertebral spine from different points of view. 

%With the rapid advances in computing and electronic imaging technology, there has been an increasing interest in developing \emph{Computer-Aided Diagnosis} (CAD) systems to improve medical services. CAD has become one of the major research subjects in medical imaging and diagnostic radiology. Although current CAD systems cannot fully replace doctors for medical detection/diagnosis in clinical practice, the results can assist doctors in providing functionalities for diagnosis, treatment, and follow-up, and for timely monitoring of disease indicators.
%
\begin{figure}[t]
\centering
\includegraphics[width=0.4\textwidth]{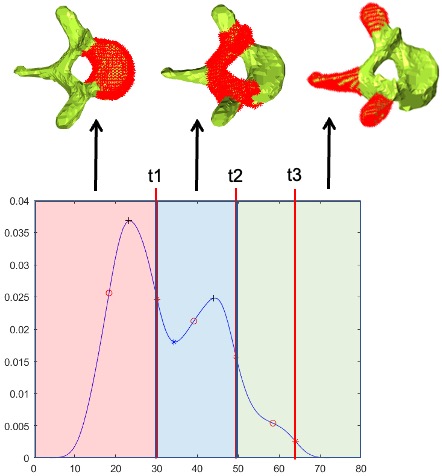}
\caption{\label{fig:probabilityDensity}Probability density curve of a vertebra and resulting segmentation associated with the inflexion points. Distances from the centroid values ($x$-axis) and probability density value ($y$-axis). The distance values corresponding to an inflexion point on the curve are considered as thresholds for the vertebral shape segmentation:~$t1$ and~$t2$ create the distinction between the three functional parts. Where~$t1$ is considered the reference value of distance for the vertebral body,~$t2$ for the vertebral arch and~$t3$ for the spinous and transverse processes region.}
\end{figure}
\subsection{Imaging and 3D modelling of the spine\label{sec:IMAGE-MODEL-SPINE}}
\paragraph*{Imaging}
For the study of the spine, medical imaging supports the evaluation of the status of the different tissues and pathological damages evolution; to this end, CT images are widely used ~\citep{bibb2014medical}. The analysed image is a greyscale image, where the tissue density is indicated by shades of grey. The \emph{Hounsfield scale} is a quantitative scale for describing radiodensity in medical CT. On the Hounsfield scale, the air is represented by a value of~$1000$ (black on the greyscale) and bone between~$700$ (cancellous bone) to~$+3000$ (dense bone) (white on the greyscale). Bones stand out clear in CT images since they are much denser than soft tissues. Different studies evaluated HU (\emph{Hounsfield Units}) values in CT for the analysis of the physio-pathological status of the subject to help in disease diagnosis~\citep{loffler2020x,zou2019use}; other works, identified a correlation between HU values and bone mineral densities~\citep{kim2019hounsfield}. This correlation is important since it allows the use of CT HU values for retrieving different information and quantitative analysis, which, in turn, reduces the need for other types of invasive images, thus reducing the overall amount of ionizing radiation for the patient. In clinical practice, as well as in different studies, the evaluation of the HU values is usually conducted on a 2D slice and in a ROI manually identified by experts~\citep{lee2013correlation},\citep{zou2019use},\citep{kim2019hounsfield}. This practice is time-consuming, error-prone, user-dependent, and does not take advantage of the modern 3D imaging techniques available.
\begin{figure}[t]
\centering
\centering
\begin{tabular}{cc}
(a)\includegraphics[height=0.20 \linewidth]{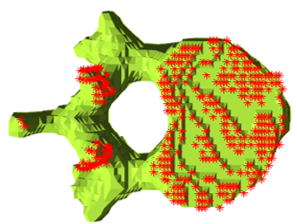}&
(b)\includegraphics[height=0.20 \linewidth]{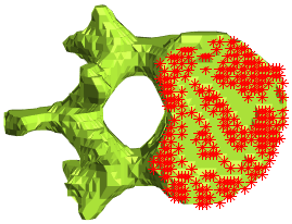}
\end{tabular}
\caption{(a) Nearest neighbor localization on consecutive vertebrae; (b) selection of the vertex belonging exclusively to the vertebral body.\label{fig:articularFacet}}
\end{figure}
\paragraph*{3D modelling}
Other works focused on the realization of 3D models to study the spine properties, considering the whole spine or the single vertebra. These models are mainly built for Finite Element analysis~\citep{barron2007generation}, e.g.,~\citep{aroeira2017three,salsabili2019simplifying,campbell2016automated,anitha2020effect} but usually are not patient-specific. Indeed, when it comes to assigning material properties of the tissue, these works rely on the state-of-the-art instead of on the single subject situation. A common technique for the development of vertebral models includes statistical shape models~\citep{castro2015statistical,campbell2016automated} or parametric model~\citep{vstern2011parametric}, which adapt to the subject-specific case with a certain level of accuracy, but usually require the construction of a prior reference.

For a deeper evaluation of the spine biomechanics, various studies focused on alignment analysis~\citep{laouissat2018classification,yeh2021deep,roussouly2011sagittal} and the study of the vertebral spine or single vertebra morphology~\citep{keller2005influence,lois20193d,casciaro2007automatic,shaw2015characterization}. The main drawback is the use of 2D images, which provide a limited visualization of the spine. 
\begin{figure}[t]
\centering
\centering
\begin{tabular}{cc}
(a)\includegraphics[height=0.20 \linewidth]{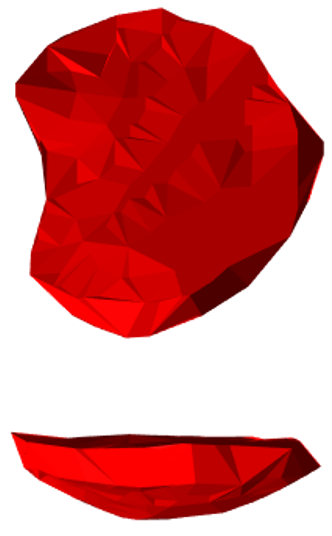}&
(b)\includegraphics[height=0.20 \linewidth]{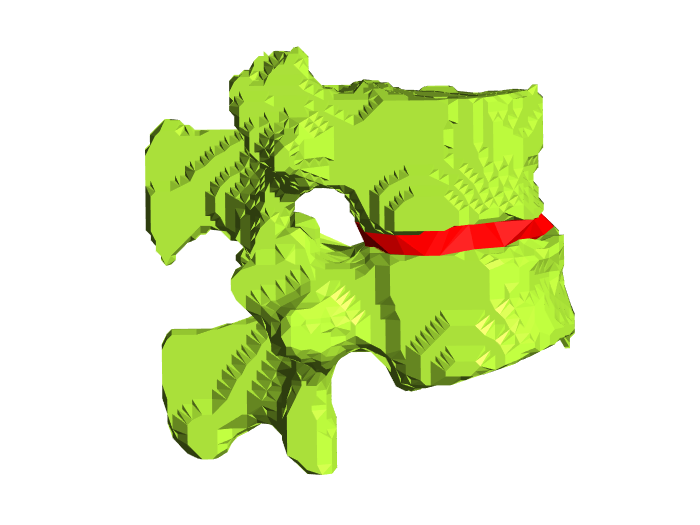}
\end{tabular}
\caption{(a) 3D surface triangulation of the inter-vertebral space; (b) inter-vertebral surface model in a moving segment of the spine.\label{fig:discSurface}}
\end{figure} 

From image analysis to modelling, the starting point is a 2D image, which shows just a view of the whole subject's anatomy. Even though rendering techniques have been developed in recent years, the common practice still considers individual slices. However, recent studies have shown that the 3D information of the vertebral spine can make a difference in early diagnosis or subject evaluation~\citep{labelle2011seeing}. In contrast to 2D image analysis, the use of 3D information allows the inclusion of asymmetries in the evaluations~\citep{barron2007generation}, improves the estimation of curvatures~\citep{lois20193d}, and encodes information on the vertebral disc status~\citep{fazzalari2001intervertebral,lois20193d}. Most of these studies starts from CT images, which have the advantage to maintain the real distances between vertebral bodies and the real curvature information~\citep{barron2007generation}. However, the works that focus on 3D modelling or imaging usually concentrate only on a specific region or motion segment and not on the complete vertebral spine~\citep{barron2007generation}, as in~\citep{lois20193d}.
\begin{figure}
\centering
\includegraphics[width=0.4\textwidth]{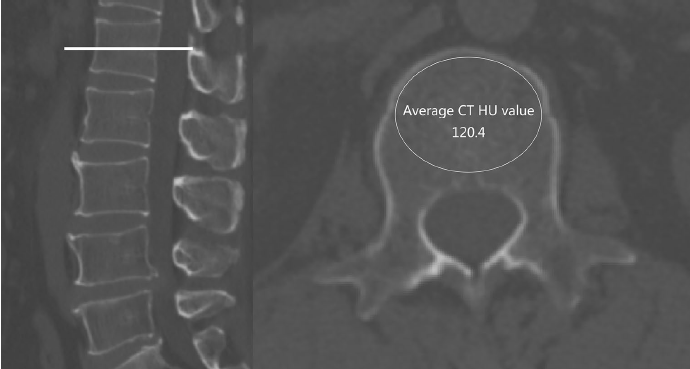} \caption{\label{fig:HU2D}Manual extraction of the mean HU value from a lumbar vertebra~\citep{zou2019use}.}
\end{figure}
\section{Proposed framework\label{Method}}
For the characterisation of the morphology and pathology of the spine, we study its 3D representation (Sect.~\ref{Sect:3DModel}) and shape segmentation (Sect.~\ref {Subsect:spineSegm}), the quantification of the intervertebral space (Sect.~\ref{sect:IntervertebralExtraction}) and the analysis of its tissues and bones (Sect.~\ref {Sect:Tissue&Bone}).
\begin{figure}
\centering
\centering
\centering
\begin{tabular}{cc}
(a)\includegraphics[width=0.35\textwidth]{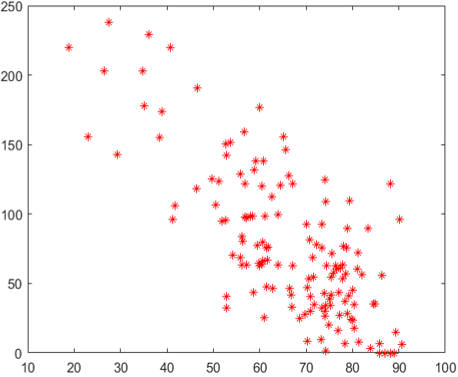}& 
(b)\includegraphics[height=0.30\textwidth]{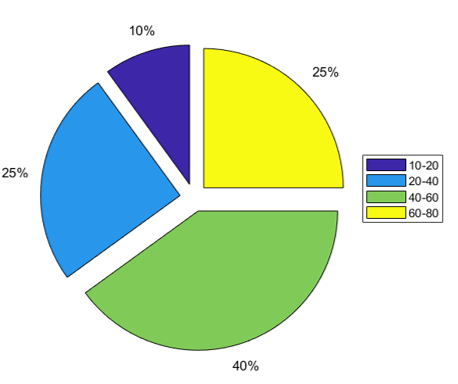}
\end{tabular}
\caption{\label{fig:densityAndAge}(a) BMD behaviour in function of subject age;~$x$-axis presents subject age,~$y$-axis BMD values. (b) Age distribution of training set.}
\end{figure}
\subsection {3D Modeling of the spine \label{Sect:3DModel}}
To analyze the spine from a 3D patient-specific perspective, we extract a 3D model from a segmented volume image. Since the spine is composed of 33 vertebrae, we extract the 3D model of each vertebra and its relation with the other vertebrae composing the spine. We assume that our input is a segmented spine data set composed of 3D segmented images, where each vertebra is labeled with a value. The labeling values go from 1 to 28 starting from the first cervical vertebra and provide the classification of the main vertebral regions: cervical, thoracic and lumbar. Thus, the single vertebra can be localized in the image space searching for the voxels that have that vertebra's specific label value. Moreover, the centroid coordinates of the vertebral bodies are provided in voxels in the image space. For more information on the data set used in this study, we refer the reader to Sect.~\ref{sect:dataset}. To obtain the 3D model of a single vertebra, we apply the~$\alpha$-shape technique, built on the image voxel centroids~\citep{edelsbrunner2010computational} and the data structure described in~\citep{paccini2020analysis}. Starting from the segmented image, we develop a 3D grid, where each element has the same dimension as the voxel in the image, and we obtain the 3D coordinates of the voxels. To build the 3D model through the~$\alpha$-shape technique, the only points required are the ones belonging to the vertebra to be reconstructed, which correspond to the centroids of the voxels belonging to the vertebra (Fig.~\ref{fig:alhashape}).
\begin{figure}[t]
\centering
\centering
\begin{tabular}{cc}
\includegraphics[width=0.13 \linewidth]{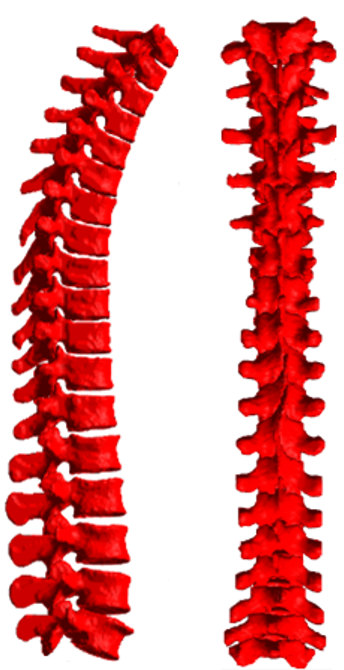}&
\includegraphics[width=0.15 \linewidth]{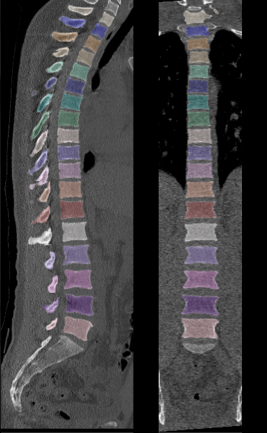}\\
(a) & (b)
\end{tabular}
\caption{\label{fig:ModeVS2D}(a) 3D model visualization from different point of view. (b) Single 2D slice visualization.}
\end{figure} 
We briefly recall that 3D~$\alpha$-shapes have been widely used in image processing and data visualization to reconstruct objects' surfaces~\citep{edelsbrunner1994three,zhou2010discriminatory}. The 3D~$\alpha$-shapes are a  generalization of the convex hull of the point: the~$\alpha$-shape of~$\mathcal{S}$ is a polytope that is neither necessarily convex nor necessarily connected. For~$\alpha = \infty$, the~$\alpha$-shape is the convex hull of~$\mathcal{S}$; as~$\alpha$ decreases, the~$\alpha$-shape shrinks. Intuitively, a piece of the polytope disappears when~$\alpha$ becomes small enough so that a sphere with radius~$\alpha$, or several such spheres, can occupy its space without enclosing any of the points of~$\mathcal{S}$. With a smaller~$\alpha$ the surface will be more detailed, for a larger~$\alpha$ the surface could lose some details.
\begin{figure}[t]
\centering
\centering
\begin{tabular}{cc}
\includegraphics[height=0.3 \linewidth]{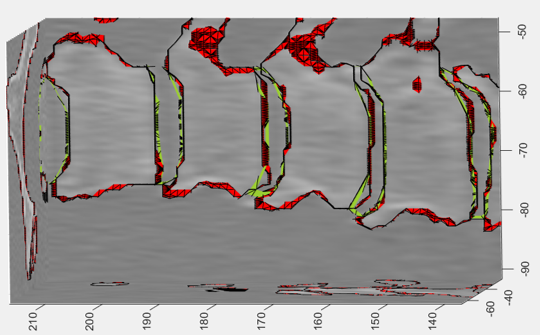}&
\end{tabular}
\caption{\label{fig:aidedVis}Volume image visualization through superimposition of the 3D patient-specific models. In CT images, soft tissue are harder to distinguish by human eyes, thus the presence of the 3D surface model helps the physician in the localization of the inter-vertebral space.}
\end{figure} 
In our work, the~$\alpha$-shape is applied to represent accurately the external surface of the vertebra, being closed, without cavities or tunnels and empty inside. For this reason, we use all the voxel's centroid belonging to the vertebra as an input set of points and considered the smallest~$\alpha$ that produces an~$\alpha$-shape enclosing all points. Then, we extract the boundary facets of the~$\alpha$-shape, to simplify the resulting triangulation and obtain only the representation of the external surface of the vertebra.  Once the~$\alpha$-shape has been extracted, the vertebra's surface is represented as a triangle mesh, which corresponds to the external surface of the volume occupied by the interested vertebra in the volume image. Basic geometric information (e.g., volume and area) are extracted from the 3D model of each vertebra. In this work, we show how the vertebral volume and surface of the specific patient behave throughout the whole spine, as well as in each segment of the spine separately.
\begin{figure}[t]
\centering
\centering
\begin{tabular}{cc}
(a)\includegraphics[height=0.20 \linewidth]{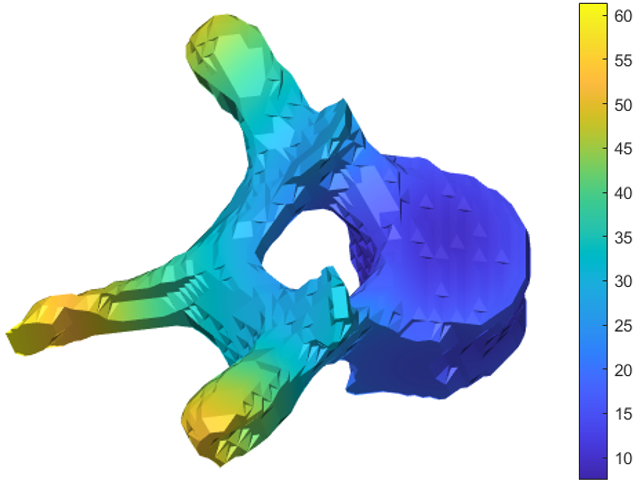}&
(b)\includegraphics[height=0.20 \linewidth]{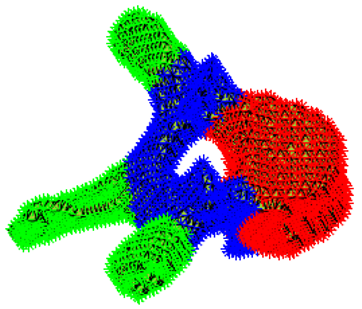}
\end{tabular}
\caption{\label{fig:segmentationVertebra}(a) Single vertebra distance distribution from the centroid. (b) Single vertebra shape segmentation.}
\end{figure} 
\subsection{Spine segmentation\label{Subsect:spineSegm}}
We propose a segmentation of the vertebral bones to highlight, the three main functional components of each vertebra (i.e., vertebral body, vertebral arch and transverse processes) from a patient-specific perspective. For the whole spine, we provide the segmentation of each vertebra and the vertebral and inter-vertebral foramina distribution.
 \begin{figure}[t]
\centering
\centering\begin{tabular}{cc}
(a)&\includegraphics[width=0.40\textwidth]{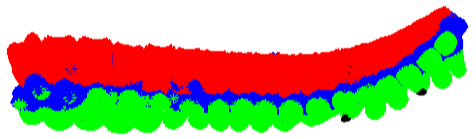}\\
(b)&\includegraphics[width=0.35\textwidth]{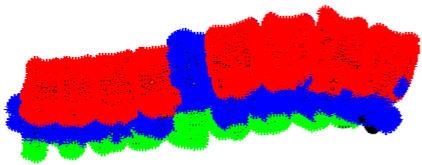}
\end{tabular}
 \caption{\label{fig:segmentationSpine} Spine segmentation result in an healthy subject (a) and in a pathological subject (b).}
\end{figure}
\paragraph{Distance distribution}
For the identification of the three functional components of each vertebra, the first step is the computation of the distance distribution of each vertex composing the vertebral surface from the centroid of the vertebral body. The centroid is provided by the data set and is identified by medical experts (Sect.~\ref{sect:dataset}). The idea is to grow a sphere centred in the centroid and to assign to each vertex of the vertebral surface a value of distance equal to the radius of the first sphere containing it. For efficiency, the sphere growth is executed by computing the Euclidean distance \mbox{$d(\mathbf{p_i},\mathbf{c}):=\|\mathbf{p_i}-\mathbf{c}\|_{2}$} of each vertex~$\mathbf{p_i}$ from the vertebral body centroid ~$\mathbf{c}$ (Fig.~\ref{fig:distanceDistribution}).
\begin{figure}[t]
\centering
\centering
\begin{tabular}{cc}
\includegraphics[height=0.25 \linewidth]{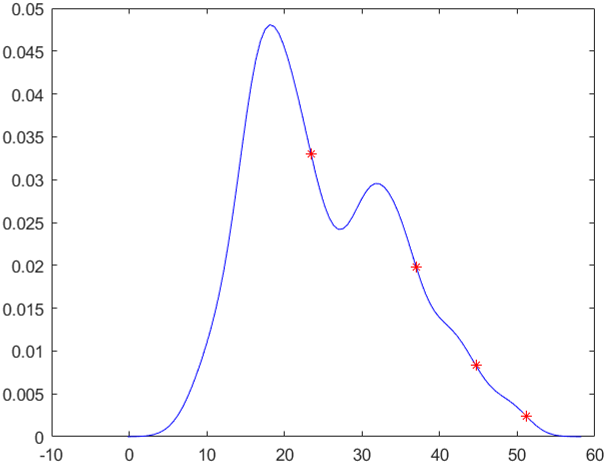}&
\includegraphics[height=0.25 \linewidth]{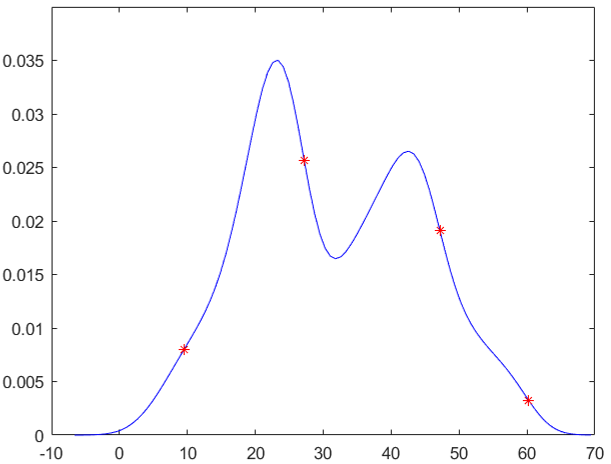}\\
(a) & (b)
\end{tabular}
 \caption{\label{fig:probDistrPatient} Difference between curve morphology and inflection point locations in healthy subjects (a) and in pathological subjects (b).}
\end{figure} 
\paragraph{Probability density}
From the distribution of distances associated with the vertices of each vertebra, we extract probability density estimates. In this way, we transform 3D information, such as the distance of the vertex from the vertebral body centroid, into 1D information. In different subjects, the same vertebra presents a similar behaviour of the probability density curve. Thus, we leverage the relevant points of such curves as thresholds to identify the different regions of the vertebral bones. In particular, the best distance thresholds for the segmentation are the inflexion points of the probability density curve. Indeed, the vertex that has a distance lower than the first inflexion point is considered a part of the vertebral body. The vertex whose distance belongs to the range between the first and the second inflexion point is considered a part of the vertebral arch. All the vertices that have a distance higher than the second inflexion point are classified as part of the transverse or spinous processes (Fig.~\ref{fig:probabilityDensity}). Once the vertebral bone has been segmented and labelled in its functional sub-parts, we perform a geometrical evaluation of each different region. In particular, we consider as a geometrical parameter the distance values corresponding to the inflexion points for each vertebra for further analysis. 
\begin{figure}[t]
\centering
\centering
\begin{tabular}{ccc}
(a)\includegraphics[width=0.45 \linewidth]{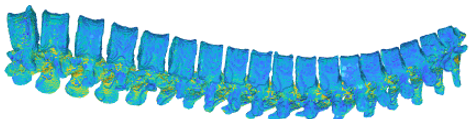}\\
(b)\includegraphics[width=0.45 \linewidth]{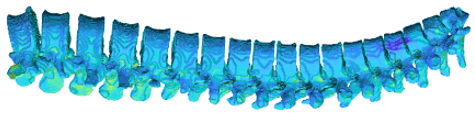}\\
(c)\includegraphics[width=0.45\linewidth]{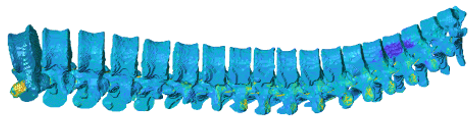}
\end{tabular}
\caption{\label{fig:mappindResult} Results of the mapping method applied to the vertebral surface model in the three different criteria: internal (a), Euclidean (b) and external (c). The difference in texture reflects the differences in tissues composition.}
\end{figure} 
\paragraph{Combined gray-level and geometric analysis}
Until this point, all the analysis and the shape segmentation have been based on the 3D surface models. Since those 3D surfaces have been obtained from a segmented image, the information related to the tissues is lost in the image segmentation. We now describe how the original CT image is used to enhance the shape segmentation and geometrical information with the volume and tissue information in a neighbourhood of the surface. To retrieve the information of the original volume, we apply the method proposed in~\citep{paccini2020analysis} and briefly described in the following.
\begin{figure}[t]
\centering
\centering
\begin{tabular}{ccc}
\includegraphics[width=0.3 \linewidth]{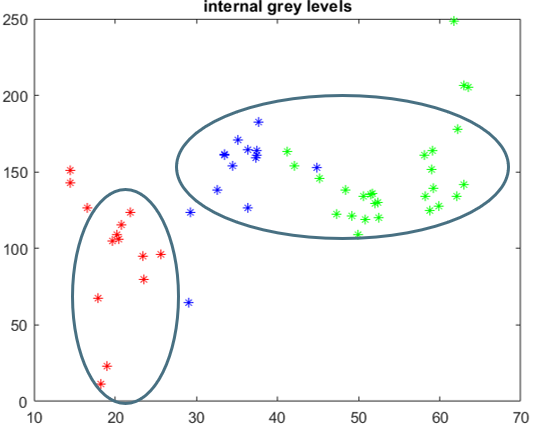}&
\includegraphics[width=0.3 \linewidth]{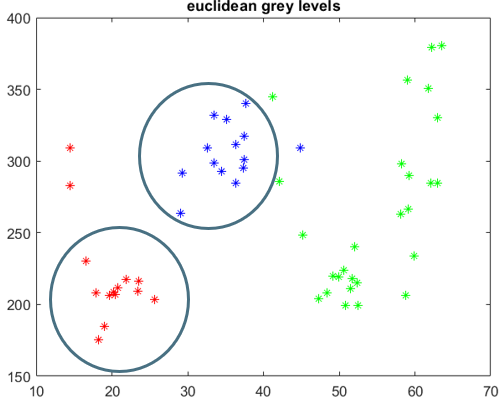}&
\includegraphics[width=0.3 \linewidth]{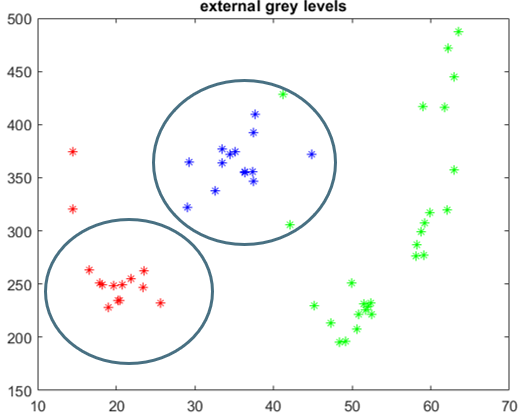}
\end{tabular}
\caption{\label{fig:comparisonResult}Comparison between tissue and geometrical information in the surface neighborhood.The~$x$-axis shows the thresholds distance values from the centroid; the~$y$-axis the mean HU values of the relative functional area localized by the thresholding, i.e. red points refer to vertebral body, blue to vertebral arch and green to spinous and transverse process.}
\end{figure}
As previously mentioned, the segmented 3D surface is a triangle mesh, the volume image is composed of a series of 2D slices and is represented as a voxel grid. The image is loaded into the 3D grid, whose elements have the same dimension as the image voxels. In this way, every grid cell has its 3D coordinates to locate it in space and each grid cell is associated with a voxel and its grey level. This data structure allows the navigation of the volume through the surface to find the correct correspondences between the volume image and the segmented 3D surface. Then, for the grey value mapping, three different criteria are defined to analyze the grey values in the proximity of the surface.
\begin{figure}[t]
\centering
\centering
\begin{tabular}{ccc}
\includegraphics[width=0.3 \linewidth]{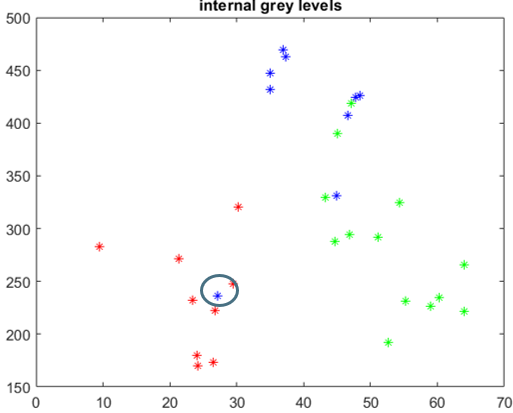}&
\includegraphics[width=0.3 \linewidth]{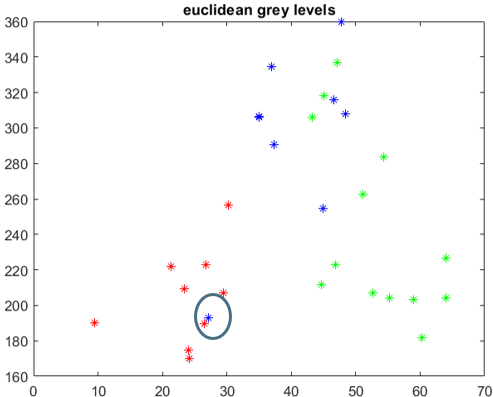}&
\includegraphics[width=0.3 \linewidth]{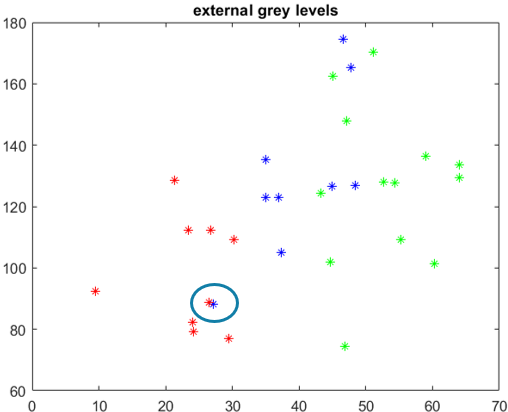}
\end{tabular}
\caption{\label{fig:comparisonResultPatient}Comparison between tissue and geometrical information in the surface neighborhood in presence of a pathological case. The~$x$-axis shows the thresholds distance values from the centroid; the~$y$-axis the mean HU values of the relative functional area localized by the thresholding, i.e. red points refer to vertebral body, blue to vertebral arch and green to spinous and transverse process.}
\end{figure}
These criteria depend on the method chosen to identify the correspondences between the surface vertices and the volume voxels. In the \textit{Euclidean mapping}, the surface vertex~$\mathbf{p}$ gets the grey-level of the voxel closest to~$\mathbf{p}$ in terms of the Euclidean distance. In the \textit{internal mapping}, the closest voxels are searched only inside the surface, that is, inside the object's volume. In the \textit{external mapping}, the closest voxels are searched only outside the surface, that is, outside the object's volume. Once every vertex has been associated with a voxel, the same voxel grey-level is associated with the correspondent vertex. The result of the grey-level mapping is a textured surface, where each vertex coordinate is associated with its specific colour, representative of the information contained in the image. This approach focuses exclusively on the regions of the volume image that are near the bone surface by mapping the grey levels in the neighbour of the surface onto the 3D shape model.
\begin{figure}[t]
\centering
\includegraphics[width=0.48\textwidth]{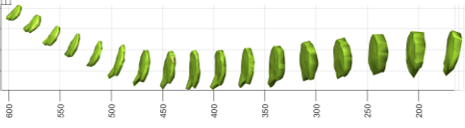}
\caption{\label{fig:IntervertExtrResult} Extraction of the intervertebral space in the whole spine.}
\end{figure}
At this point, we have a 3D surface textured with the original volume grey levels, meaning that each vertex of the triangulation is associated with a value extracted from the CT. Indeed, we can link the information related to the segmentation of the 3D surface into the three functional parts, with the texture information obtained with the texture mapping method. In particular, we correlate the geometric thresholds localizing the vertebral region (i.e., the distance thresholds corresponding to the inflexion point on the probability distribution curve:~$t1$,~$t2$ and~$t3$ in Fig.~\ref{fig:probabilityDensity}), with the mean grey value of the region in all three mapping method criteria. In this way, we characterize the spinal district considering both tissue and geometric information and we evaluate which of the mapping criteria is best suited for this characterization.
\begin{figure}[t]
\centering
\includegraphics[width=0.4\textwidth]{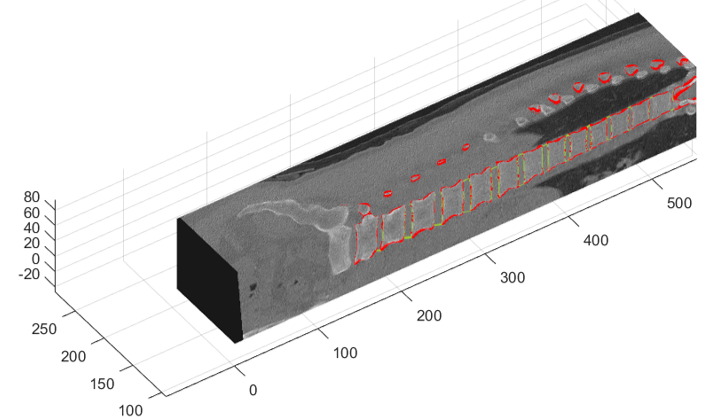}
\caption{\label{fig:IntervertVolume} Combined visualization: 3D volume with segmented surfaces.}
\end{figure}
\begin{figure}[t]
\centering
\includegraphics[width=0.40\textwidth]{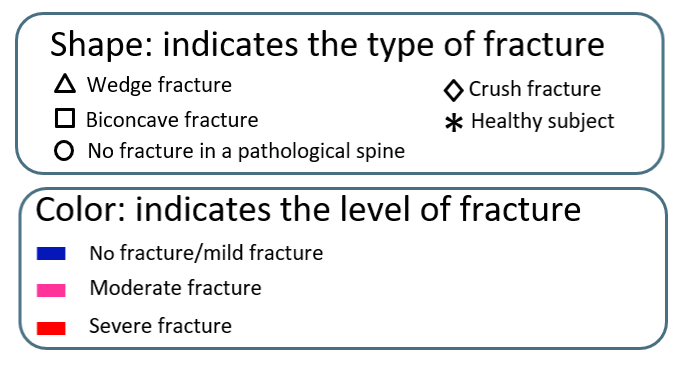}
\caption{\label{fig:legend} Legend for the interpretation of the subject comparison graphs in Figs.~\ref{fig:IntervertGeometric}-\ref{fig:VertBodyCombined}.}
\end{figure}
\subsection{Intervertebral space\label{sect:IntervertebralExtraction}}
The intervertebral disc is the structure that bonds adjacent vertebral bodies. There are 23 inter-vertebral discs, where the first locates between C2 and C3 and the last disc is at the lumbosacral junction. A thin plate of hyaline cartilage covers the upper and lower surfaces of vertebral bodies interested in the junction and the disc is interposed between these hyaline cartilage layers. The discs located in the thoracic region are thinner than the ones in the lumbar region. Moreover, the discs help the lumbar part of the vertebral column to assume the pronounced physiological lordosis curve since they are thicker anteriorly and narrow posteriorly.
\begin{figure}[t]
\centering
\centering
\begin{tabular}{cc}
\includegraphics[width=0.35 \linewidth]{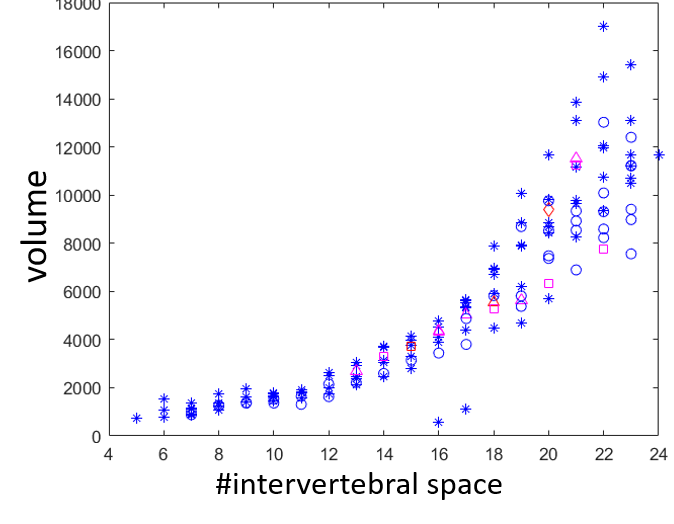}&
\includegraphics[width=0.35 \linewidth]{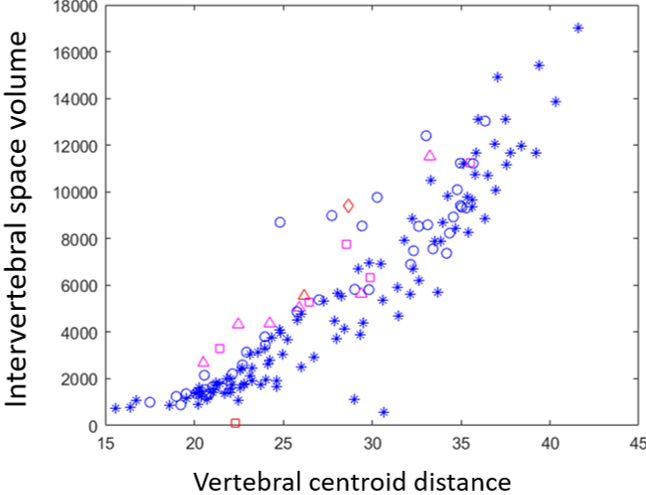}\\
\end{tabular} 
\caption{\label{fig:IntervertGeometric} With reference to the legend in Fig.~\ref{fig:legend}, analysis of geometrical parameters of the intervertebral space.}
\end{figure}
Each intervertebral disc is made up of two structures: the annulus fibrosus, which is an outer lamellated (multi-layered) fibrous ring, and the nucleus pulpous, which is the gelatinous inner zone. The twofold composition of the disc is required to make it an efficient shock-absorber. The annulus is structured to deal with torsional and shear stresses to which it is subjected, while the nucleus pulposus functions as a hydrated gel and possesses compressibility.
\begin{figure}
\centering
\includegraphics[width=0.70\textwidth]{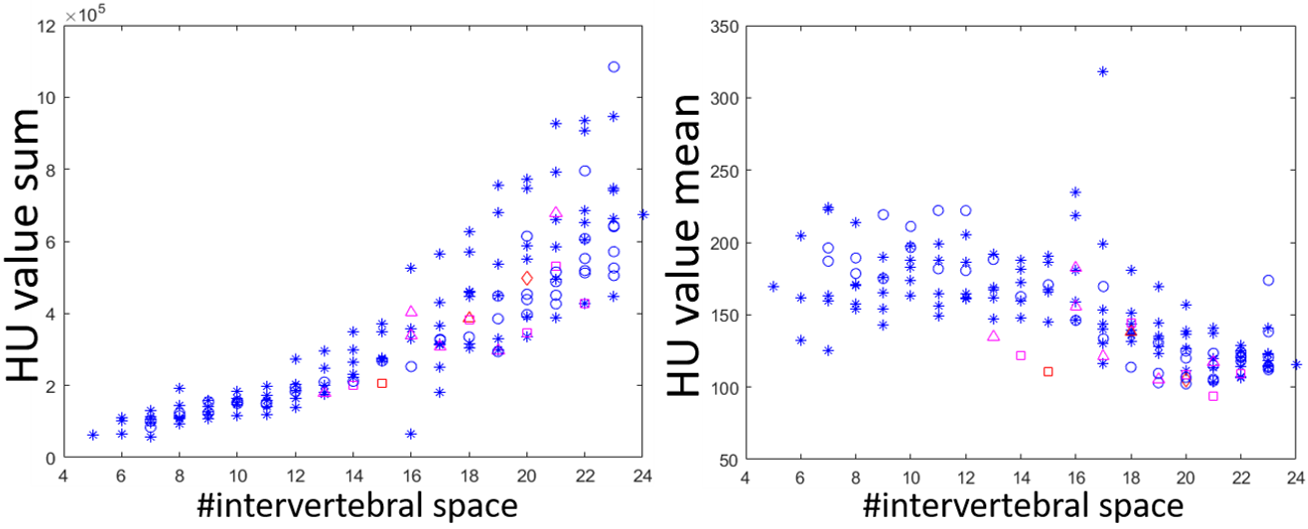}
\caption{\label{fig:IntervertTissue} Tissue information evaluation in the inter-vertebral space.}
\end{figure}
Through the CT images composing the considered data set, we are not able to distinguish and extract the various component of the inter-vertebral junction and the intervertebral disc. Nevertheless, we localize the intervertebral space between two consecutive vertebrae leveraging the 3D surface models. This information helps in the 3D characterization of the district other than facilitating the localization of the intervertebral space in the 3D CT, which does not highlight soft tissue as well as it does with bony structures. 
\begin{figure}[t]
\centering
\includegraphics[width=0.70\textwidth]{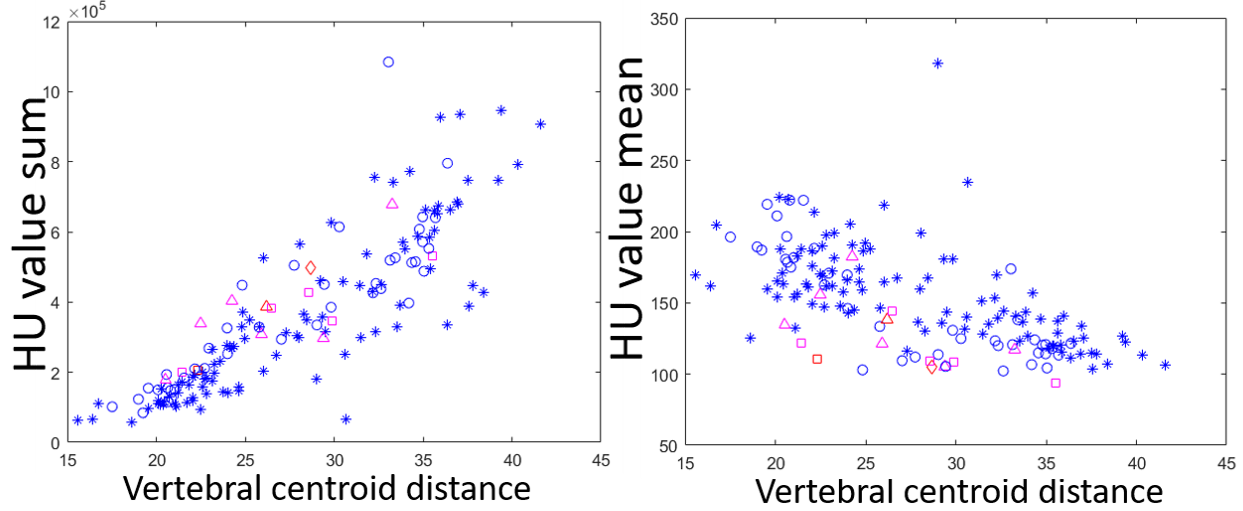}
\caption{\label{fig:IntervertCombined} Evaluated tissue information in the inter-vertebral space in relation with geometric information.}
\end{figure}
The intervertebral space involves only the portion of space between two consecutive vertebral bodies. To extract it from a 3D patient-specific perspective, we use the 3D surface triangulation obtained from the segmented image, searching the vertebral body portion that faces the subsequent or prior vertebra. To extract and locate each inter-vertebral space throughout the whole vertebral spine, we consider two consecutive vertebrae at a time. For each couple of vertebrae, we identify the vertices belonging to one vertebra and nearest to the other vertebra. To this end, we apply a~$k$d-tree approach where one of the vertebrae is used to construct the model and the vertices of the second vertebra are query points. We repeat the reverse operation to obtain the nearest neighbour vertices in both the upper and lower vertebra (Fig.~\ref{fig:articularFacet}(a)).

Since this operation identifies the vertices in the vertebral body, but also the vertices in the vertebral arch and the transverse and spinous processes, it is necessary to select only the vertices belonging to the vertebral body. To this end, we apply the same method and the same thresholds applied to the vertebral surface segmentation (Sect.~\ref{Subsect:spineSegm}). Ideally, the remaining vertices are those facing the upper/lower vertebra and included in a sphere centred in the vertebral body centroid, with a ray equal to the distance at which the probability density curve (obtained from the vertex distance distribution) presents the first inflexion point (Fig.~\ref{fig:articularFacet}(b)). In this way, we obtain a point cloud that surrounds the intervertebral space. Using this point cloud, we build an~$\alpha$-shape and obtain a 3D triangulation of the intervertebral space, thus the 3D surface model that we were looking for (Fig.~\ref{fig:discSurface}).  

Analogously to the 3D surface model of the vertebrae, the 3D triangulation of the intervertebral space allows us to easily compute geometrical parameters related to the surface. The advantage of using a 3D surface is that the geometrical parameters computed will evaluate the intervertebral space by taking into consideration its 3D shape. In particular, we consider the volume of the intervertebral space for further analysis. 
\subsection{Tissue and bone analysis\label{Sect:Tissue&Bone}}
Voxel grey levels of CT images are expressed in Hounsfield Units (HU)~\citep{Hounsfie99:online}. It has been proven that there is a correlation between the T values of the DEXA measurement and the HU of the same vertebral body. The DXA-based T-score is used for the diagnosis of osteoporosis. Thus, the significant correlation of HU values with DXA T-scores means that patients' HU values can indicate to physicians the presence of osteoporosis~\citep{lee2013correlation}. For this reason, the analysis of the HU values during the evaluation of the vertebral spine is really important and helps to predict the curse of a disease, such as osteoporosis or to plan the surgery. In clinical practice, the evaluation of HU vertebral values is still performed manually and through the observation of a single 2D CT slice~\citep{zou2019use},\citep{kim2019hounsfield} (Fig.~\ref{fig:HU2D}). 

We now describe how, leveraging the 3D surface model of the spine obtained from the segmented volume image (Sect.~\ref{Sect:3DModel}), we evaluate the HU values on each vertebra in an automatic way and from a 3D patient-specific perspective. We are provided with the centroid of the vertebral body of each vertebra that appears in the CT image (Sect.~\ref{sect:dataset}). Moreover, from the segmented image, we can extract the 3D surface model of every single vertebra represented (Sect.~\ref{Sect:3DModel}). Our goal is to identify a 3D ROI that will be the area where the HU values will be evaluated. In particular, once identified the ROI we consider all the voxels contained in it and we compute the average and sum of the HU values presented by those voxels. 

To build the 3D ROI, we grow a sphere centred in the centroid of the vertebral body until it reaches the external 3D surface of the vertebra; thus, the sphere stops growing when it first touches the boundary of the vertebral surface. The sphere, as well as all the other 3D surfaces, is built through~$\alpha$-shape techniques and, is represented as a 3D triangulation. The initial ray considered for the sphere is equal to the extension of a voxel diagonal, then for each step, the ray is increased by one diagonal dimension at a time. Superimposing the 3D models of the sphere, and the vertebra to the 3D original CT (loaded on the 3D grid described in Sect.~\ref{Sect:3DModel}), we can localize all the voxels that fall inside the sphere. In particular, we consider a voxel as belonging inside the sphere if its centroid is included in the sphere (i.e., if the distance of the vertebra centroid from the voxel centroid is smaller than the sphere ray). Given~$r$ the radius of the maximum sphere~$\mathcal{S}$ contained in the vertebral body,~$\mathbf{v_i}$ a centroid of the voxels belonging to the vertebra and~$\mathbf{c}$ the centroid of the vertebral body,~$\mathbf{v_i} \in\mathcal{S}\Leftrightarrow  \mbox{$d(\mathbf{v_i},\mathbf{c})$} \leq r$, where the distance of each voxel from the vertebral body centroid is~$\mbox{$d(\mathbf{v_i},\mathbf{c})$} = \|\mathbf{v_i}-\mathbf{c}\|_{2}$.

Once all the voxels inside the ROI are identified, we can compute the mean value of the grey levels (HU values) that those voxels preset. To further evaluate the bone tissue we also compute the sum of the HU values presented by the ROI voxels. At the end of this procedure, we have both information on the tissue inside the vertebral volume and geometric information regarding the vertebral body indicated by the maximum ray obtained by growing the sphere. To improve the spine characterization we can combine this information to provide an evaluation that considers all the possible aspects and leverages all the information retrievable from the data set. 

\section{Results and discussion\label{sec:RESULTS}}
We discuss the experimental results of the proposed approach (Sect.~\ref{sec:EXPERIMENTS}) on benchmarks of the spine district (Sect.~\ref{sect:dataset}), thus showing the main properties of the proposed framework in terms of characterisation of the morphology and pathologies of the spine.

\subsection{Data set \label{sect:dataset}}
The CT data used in this work are a sub-part of the \emph{Large Scale Vertebrae Segmentation challenge} (VerSe)~\citep{SEKUBOYINA2021102166}, which provides a common benchmark for current and future spine-processing algorithms~\citep{SEKUBOYINA2021102166}. It consists of multi-detector, spine CT (MDCT) scans with vertebral-level (3D centroids) and voxel-level annotations (segmentation masks). The data are multi-site acquired using multiple CT scanners, present a variety of FoVs (including cervical, thoracolumbar and cervical-thoracolumbar scans), and a mix of sagittal and isotropic reformations, and cases with vertebral fractures, metallic implants, and foreign materials~\citep{SEKUBOYINA2021102166}. % Fig. 1 illustrates this variability in the VerSe dataset. Refer to Löffler et al. (2020b); Liebl et al. (2021) for further details on the data composition.

The annotations in the data set are twofold and present 3D coordinate locations of the vertebral centroids and voxel-level labels as segmentation masks. Twenty-six vertebrae (C1 to L5, and the transitional T13 and L6) are annotated with labels from 1 to 24, along with labels 25 and 28 for L6 and T13, respectively.
%
%\begin{itemize}
%    \item~$1-7$ refer to cervical spine: C1-C7
%    \item~$8-19$ refer to thoracic spine: T1-T12
%    \item~$20-25$ refer to lumbar spine: L1-L6
%    \item~$26$ refers to sacrum - not labeled in this data set
%    \item~$27$ refers to cocygis - not labeled in this data set
%    \item~$28$ refers to additional 13th thoracic vertebra, T13
%\end{itemize}
%
The vertebrae that are partially visible at the top or bottom of the scan are not annotated. This CT data set contains 160 image series of 141 patients including segmentation masks of 1725 fully visualized vertebrae; it is split into a training data set (80 image series, 862 vertebrae), a public validation data set (40 image series, 434 vertebrae), and a test data set (40 image series, 429 vertebrae)~\citep{doi:10.1148/ryai.2020190138}. The metadata includes annotation of vertebral fractures obtained through the Gentant's method, the indication of foreign material, and the measurement of lumbar bone mineral density per patient age (Fig.~\ref{fig:densityAndAge}(a)). For our purposes, we used only one type of CT scan in the training, to be able to compare the results obtained in the tissue evaluation set. Fig.~\ref{fig:densityAndAge}(b) shows the age distribution of the subjects considered in our work. %it is possible to observe the behaviour of such bone mineral density with the age of the subject. The vertebral fractures annotation works as follows: mild fractures consist of a height loss~$\geq 20 \%~$ and~$< 25\%$, a moderate fracture of a height loss of~$\geq 25\%$ and~$< 40\%$, and a severe fracture of a height loss~$\geq 40\%$. The type of fracture are categorized as \textit{wedge} (anterior height loss most prominent), \textit{biconcave} (central height loss most prominent with almost equal anterior and posterior height loss), or \textit{crush} (posterior height loss most prominent or uniform height loss including the posterior vertebral wall)~\citep{doi:10.1148/ryai.2020190138}. Patient characteristics (sex, age, BMD, contrast media applied, the scanner used) were not significantly different between training and both test data sets. The data are hosted at the open science framework (\url{https://osf.io/nqjyw/}).
\subsection{Experimental results\label{sec:EXPERIMENTS}}
We now illustrate the results obtained with all the methods described in Sect.~\ref{Method} and all the analyses that we performed with the geometric and tissue-related parameters extracted from both the original CT and the 3D surface model. More precisely, we discuss the 3D modelling and enhanced visualization (Sect.~\ref{sec:ENHANCED-VISUALIZATION}), the spine segmentation and combined evaluation (Sect.~\ref{sec:SPINE-SEGMENTATION}), the intervertebral space extraction (Sect.~\ref{sec:INTER-SPACE}), the tissue and bone analysis (Sect.~\ref{sec:TISSUE-BONE}).
\subsubsection{3D modelling and enhanced visualization\label{sec:ENHANCED-VISUALIZATION}}
The use of 2D CT slice evaluation gives a partial perspective on the subject, while our 3D patient-specific surface model allows us to observe the vertebral spine in every direction (Fig.~\ref{fig:ModeVS2D}). During the evaluation of the morphological characteristics of the patient, a complete visual experience is crucial. The possibility to extract every single vertebra helps also to focus the analysis on a specific portion of the vertebral spine without loss in quality of the representation. Since the 3D models are extracted directly from the volume image, the 3D surfaces of the vertebrae will guide the visualization of the image itself by localizing the boundary of the bones inside the volume (Fig.~\ref{fig:aidedVis}). 
\subsubsection{Spine segmentation and combined evaluation\label{sec:SPINE-SEGMENTATION}}
Fig.~\ref{fig:segmentationVertebra} shows the results of the thresholding of the distance distribution which leads to the shape segmentation described in Sec.~\ref{Subsect:spineSegm}. The red part of the surface represents the vertebral body, the blue one the vertebral arch and the green one the transverse and spinous processes. Applying the segmentation to all the vertebrae represented in the image, we obtain the segmentation of the whole vertebral spine (Fig.~\ref{fig:segmentationSpine}(a)).
On a pathological subject, the shape and morphology of the vertebra can change drastically as a result of different processes that damage the tissue, thus the 3D surface models highlight such morphological changes. Applying our segmentation to a pathological case that presented a vertebral fracture due to osteoporosis, the fractured vertebra was identified among the others by a clear change in the segmentation result (Fig.~\ref{fig:segmentationSpine}(b)). Indeed, a change in the distance from the vertebral centroid distribution, in turn, modifies the vertebra's probability distribution curve as well as the position of the inflexion points. 
Fig.~\ref{fig:probDistrPatient} highlights how the inflexion points' location on the probability distribution has changed due to the osteoporosis processes that brought the vertebral fracture.

The application of the grey-level mapping algorithm to the vertebral surface model produces a 3D textured representation of the spine, where the texture corresponds to the HU values of the original 3D image. Depending on the mapping criteria we can investigate the tissues inside, across or outside the vertebral surface (Fig.~\ref{fig:mappindResult}). To combine the geometrical information retrieved from the 3D model and the tissue information given by the grey-level mapping, we evaluate the mean HU values of the vertebra's functional region according to the distance values of its inflexion point on the probability distribution ($t1$,~$t2$ and~$t3$). In this case, the geometrical information corresponds to the threshold used to segment the vertebral shape model in the three functional parts, while the tissue information is the mean HU value of the correspondent vertebral functional component. Fig.~\ref{fig:comparisonResult} shows the results obtained in the comparison of geometrical and tissue information. In each different mapping criteria, the graphs show various clusters. In the internal mapping case, the clusters could be related to a varying percentage of cortical bone in the different vertebral components. Indeed, the body is mainly composed of cancellous bone, which is the spongy type of osseous tissue. This cancellous bone is covered by a thin coating of cortical bone (or compact bone), which is the hard and dense type of bony tissue. The vertebral arch and processes have thicker coverings of cortical bone.

The clusters identified in the external mapping results can be related to the tissue surrounding the vertebra from the outside. The vertebral body confines by cartilaginous tissues (intervertebral disc) while the other parts of the vertebral border with ligaments and tendons (e.g. vertebral arch with the foramen). The slightly different values of HU between the Euclidean and external mapping results can be explained by the ligament insertions distribution on the vertebral arch and body. Indeed, the bond between two vertebral bodies is reinforced anteriorly by the anterior longitudinal ligament and posteriorly by the posterior longitudinal ligament, known also as the anterior and posterior spinal ligaments, respectively.

In the remaining parts of the vertebrae, there is no intervertebral disc but, in addition to the facet joints that connect adjacent vertebral arches, there are a few strong and bio-mechanically important ligaments that run between adjacent vertebral arches, which influence the stability of the vertebral column~\citep{MAHADEVAN2018327}. These ligaments fuse with the anterior and posterior parts of the circumference of the disc. Indeed, the Euclidean mapping, which considers the voxel crossed by the surface, is influenced by both the aspects just exposed.
In the case of healthy subjects, the comparison of heterogeneous information can differentiate the functional components of the vertebra, thus providing an insight into the tissue status. Indeed, the distribution of the points in the combined graph could represent, from a clinical point of view, the patient-specific insight into the healthiness of the patient. Comparing the result with the combined analysis of a pathological subject (Fig.~\ref{fig:comparisonResultPatient}), the point in the graph related to the vertebral body of the patient (which is the functional part mostly affected in this case) is an outlier of the results of the healthy subject. Both the graphs related to the internal and external mapping can differentiate between healthy and pathological subjects; in fact, once the osteoporosis processes damage and erode the bone tissues, the cartilaginous tissues are involved and suffer as well. This means that the tissue changes are highlighted not only by exploring the volume inside the surface but also by analyzing the volume in the region outside the bone surface. 
Thus, these representations of combined information are amble to highlight pathological cases or vertebrae that differentiate from the classical health distribution, making it a useful tool for a personalized and efficient analysis.
\subsubsection{Intervertebral space extraction\label{sec:INTER-SPACE}}
The results (Fig.~\ref{fig:IntervertExtrResult}) of the extraction of the intervertebral space from the 3D surface models (Sect.~\ref{sect:IntervertebralExtraction}) show how the inter-vertebral spaces interact with the vertebral models of the whole spine captured in the image. The advantage of having the inter-vertebral space boundary is the localization of the different structures in the volume (Fig.~\ref{fig:IntervertVolume}). This augmented visualization helps physicians in the development of accurate patient-specific evaluations. Moreover, it can be a guide in the preoperative planning phase for surgical interventions. In all the following graphs, which include a comparison between healthy and pathological subjects, the reader can refer to the legend in Fig.~\ref{fig:legend}.

The purely geometrical parameters retrieved from the inter-vertebral space extraction are in line with what is expected from the anatomy. Fig.~\ref{fig:IntervertGeometric} shows the volume occupied by inter-vertebral spaces of healthy and pathological subjects throughout the whole spine. The volume globally increases from the cervical to the lumbar segment of the spine. Moreover, we evaluated the inter-vertebral volume in function of the distance between the correspondent two vertebrae's centroids. In Fig.~\ref{fig:IntervertGeometric}, the patient's inter-vertebral volume tends to be higher than the healthy inter-vertebral volume in correspondence of the same centroid distance. This result can be related to the erosive processes that start in the more external region of the bone, thus allowing the intervertebral region to occupy a larger volume.

Considering the tissue-related information, we compute the mean HU value presented by the voxels of the intervertebral space, and also the sum of such HU values (Fig.~\ref{fig:IntervertTissue}). For both healthy and pathological subjects, we show the values obtained for each intervertebral space. To combine geometrical and tissue information, we also evaluated the mean HU value and the sum of the HU value in the intervertebral space, in the function of the distance between the centroid of the relative two vertebrae. In Fig.~\ref{fig:IntervertCombined}, the mean HU value is more significant in distinguishing pathological cases from healthy subjects. 
\subsubsection{Tissue and bone analysis\label{sec:TISSUE-BONE}}
\begin{figure}[t]
\centering
\includegraphics[width=0.65\textwidth]{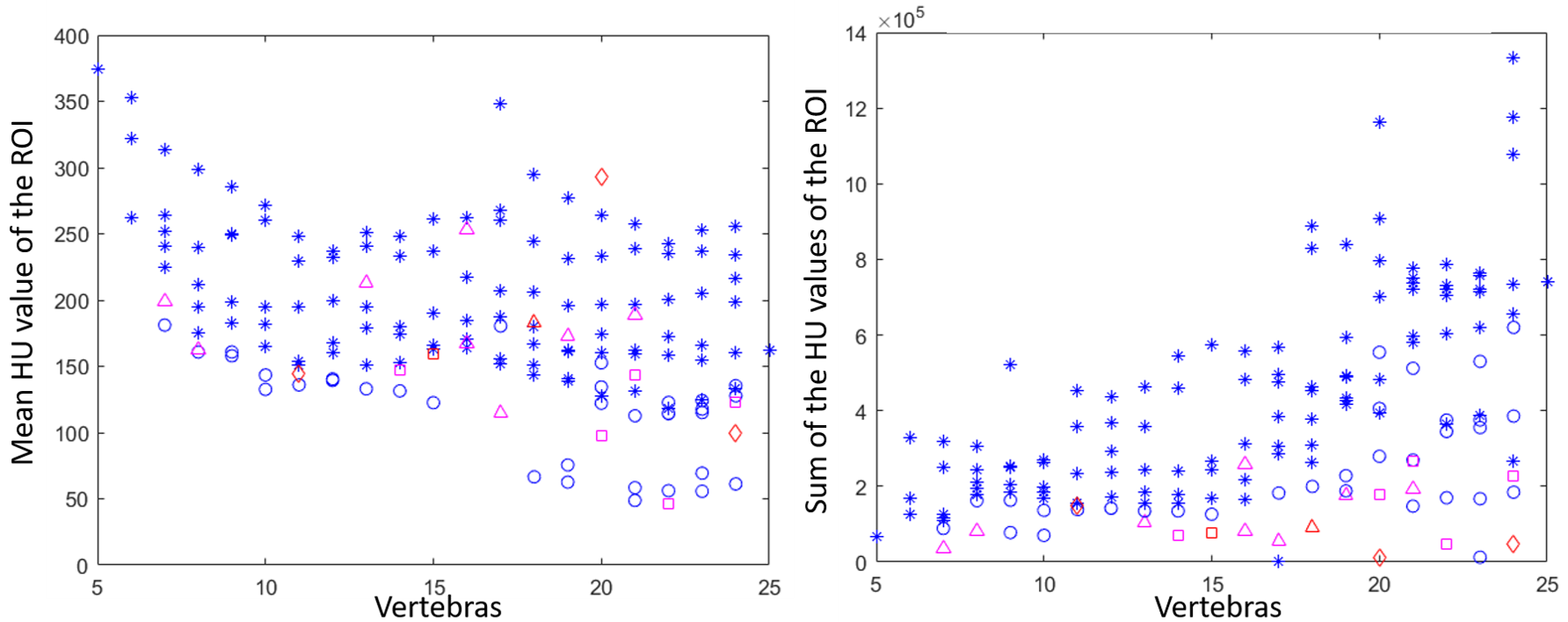}
\caption{\label{fig:VertBodyTissue} Behavious of the vertebral body tissue.}
\end{figure}

Analyzing the grey-levels mapping, we obtain a characterization of the spine in correspondence with the surface of the vertebral bones. To provide a more complete characterization, it is necessary to consider the volume inside each vertebra. From a tissue point of view, this evaluation is commonly performed by considering a manually selected ROI on a 2D CT slice and computing the average HU value. Our method improves this practice since it considers a 3D ROI automatically computed and centred in the barycenter of the vertebral body. From the analysis of the inner tissue of the vertebral body (Fig.~\ref{fig:VertBodyTissue}), we observe that the patient data are below the healthy subject ones in the graph, especially when considering the sum of the HU values inside the ROI. Indeed, the tissue information is important for the characterization of the tissue status and our evaluation helps in the differentiation between healthy and pathological subjects.

\begin{figure}[t]
\centering
\includegraphics[width=0.70\textwidth]{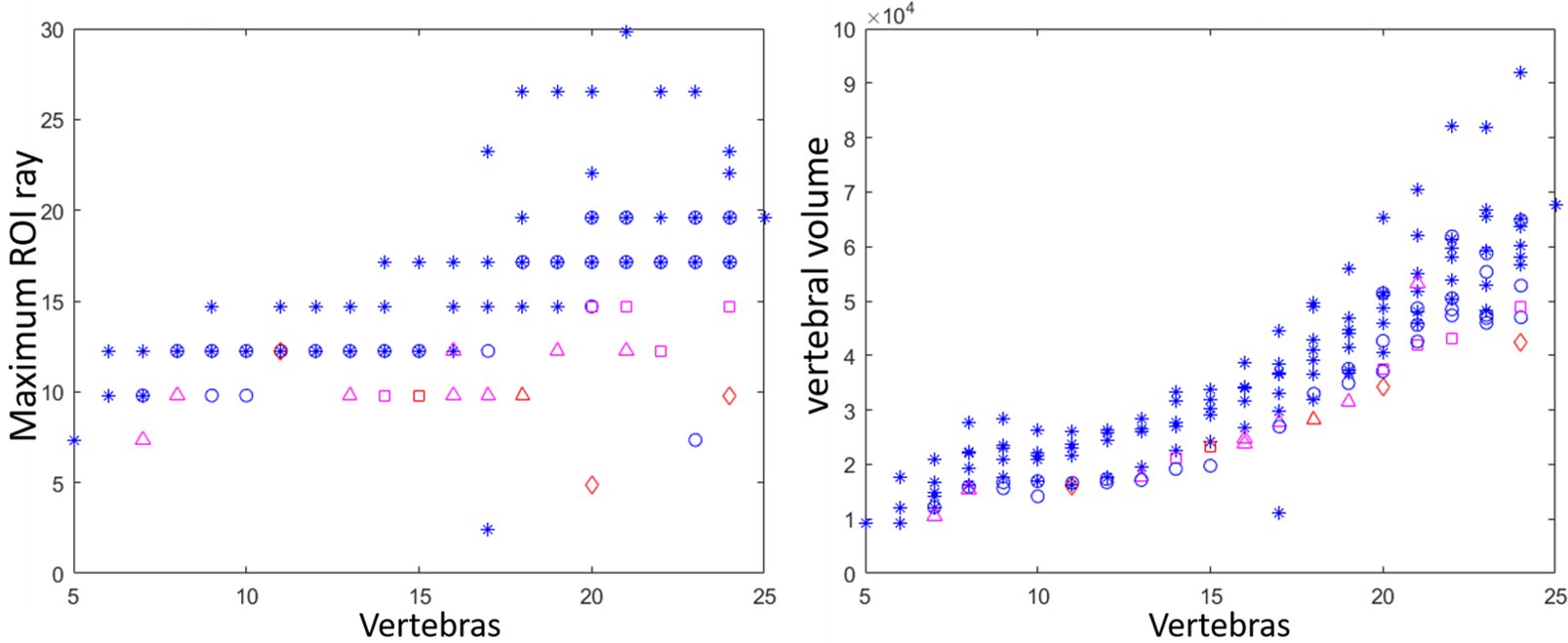}
\caption{\label{fig:VertBodyGeom} Vertebral body geometrical parameters behavior.}
\end{figure}

Since we developed 3D models for both the vertebrae and the different ROI that we locate inside the vertebral body, we consider how the ray of the spherical ROI behaves in the different segments of the vertebra (Fig.~\ref{fig:VertBodyGeom}(a)) and also how the whole vertebral volume changes moving from cervical to lumbar vertebrae (Fig.~\ref{fig:VertBodyGeom}(b)). Both the ROI radius and the vertebral volume grow going from the cervical to the lumbar segment, as expected from anatomical studies~\citep{MAHADEVAN2018327}. For the ROI ray graph, the patient data are below the healthy subject's data, which is coherent with hypotheses of bone erosion and morphological changes due to the pathology. Then, the vertebral body suffers the most from the pathology evolution, between healthy and pathological subjects. The whole volume considers also the other regions of the vertebra, thus the distinction is less clear. 
\begin{figure}[t]
\centering
\includegraphics[width=0.70\textwidth]{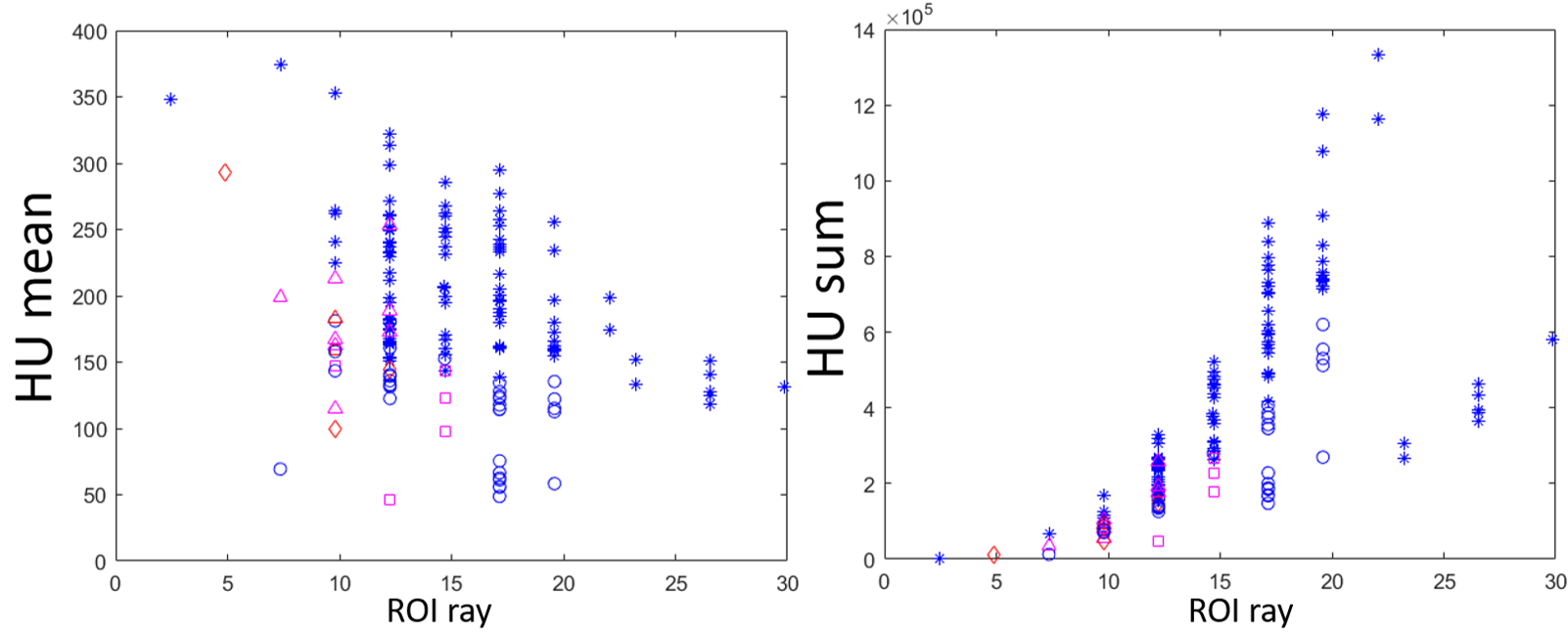} \caption{\label{fig:VertBodyCombined} Vertebral body geometrical and tissue information.}
\end{figure}
Finally, we compare the geometrical and tissue information. Fig.~\ref{fig:VertBodyCombined}(a) shows the behaviour of the HU mean in the ROI region for a given ROI ray, while Fig.~\ref{fig:VertBodyCombined}(b), provides the HU sum in the ROI for a given ROI ray. The best way to distinguish a patient from a healthy subject, in this case, is through the HU mean in the function of the ROI ray. Indeed, the ROI extension and the values of the HU mean are lower for healthy subjects. 

\section{Discussion and future work\label{sec:DISCUSSION}}
We provide a framework for the patient-specific characterization of the vertebral spine, which works on 3D surface models and permits the evaluation of the health status of each patient by analyzing the spine in terms of anatomical components, functional regions and tissue status. In this way, the visualization of the single bone and the whole district is improved, and the quantitative analysis relies on the information collected considering every direction in space, which is not possible with 2D images.

The framework brings and elaborates information based on the analysis of geometrical features, such as the volume of each vertebra or the radius of the ROI contained in the vertebral body and the tissue information, retrieved from the original CT image of the single patient. This information is obtained by navigating the volume of the image and taking as references the surface of the vertebral models and the barycenter of the vertebral body. The first ones are used for the mapping algorithm and help to visualise the volume in the neighbour of the vertebral boundaries, considering the outside, the inside, or the voxels crossed by the surface. The barycenters, instead, help in the construction of the spherical ROI that permits the analysis of the tissue inside the vertebral body. In this way, all the interesting areas of the image are considered, giving a complete characterization of the patient from the tissue point of view. According to the experimental validation, both the geometrical and the tissue analysis are coherent with what is expected from the anatomy and can help in the distinction between healthy and pathological subjects. Indeed, they could support physicians in the evaluation of the subject's status. All the data representation proposed showed the potential to distinguish between healthy and pathological subjects from different perspectives.

The main limitation of our work is the need for a segmented 3D image since the 3D model required for the development of the framework is extracted from the segmented 3D image. However, leveraging deep learning methods, the segmentation can be obtained automatically with good accuracy and reviewed by experts, thus reducing drastically the time required by this operation. Moreover, the labeling of the vertebrae and the segmentation are essential steps in processing spine data, and represent the starting point for further analyses, such as the detection and grading of fractures, calculation of bone mineral density but also analysis of spinal shape, curvature, and deformity, such as scoliosis~\citep{liebl2021computed}.

Another limitation regards the subject's posture. In our study, the CT images show the patient in a supine position with an effect of gravity lower than the standing posture. The 2D CTs used in clinical practice have the great advantage to capture the standing patient, and thus they show the spine behaviour under different loads. However, all the methods developed are general enough to be applied to different imaging modalities and to further characterize the standing modality of the subjects. 

Future work will be focused on improving the spine characterizations while keeping a clinical application perspective, further works on this topic will include the localization of posture landmarks, such as spinous processes, and their visualization on the skin. This work could provide an external reference point for physicians and surgeons during their posture evaluations. The visualization of the landmarks on the skin of the patient could lead also to integration with 3D sensors' data for posture analysis and rehabilitation applications. Furthermore, to extend the geometrical and tissue evaluation performed in our work, an extension to the analysis of the vertebral foramina would be interesting. Finally, the characterization could be improved by considering further geometrical parameters and image pre-processing or enhancement techniques.

\paragraph{Acknowledgements}
Work supported by Regione Liguria, Fondo Sociale Europeo (FSE) 2014-2020

\bibliographystyle{unsrtnat}
\bibliography{VertebralSpine}  %%% Uncomment this line and comment out the ``thebibliography'' section below to use the external .bib file (using bibtex) .

\begin{thebibliography}{31}
\providecommand{\natexlab}[1]{#1}
\providecommand{\url}[1]{\texttt{#1}}
\expandafter\ifx\csname urlstyle\endcsname\relax
  \providecommand{\doi}[1]{doi: #1}\else
  \providecommand{\doi}{doi: \begingroup \urlstyle{rm}\Url}\fi

\bibitem[Paccini et~al.(2020)Paccini, Patan{\'e}, and
  Spagnuolo]{paccini2020analysis}
Martina Paccini, Giuseppe Patan{\'e}, and Michela Spagnuolo.
\newblock Analysis of {{3D}} segmented anatomical districts through grey-levels
  mapping.
\newblock \emph{Computers $\&$ Graphics}, 91:\penalty0 179--188, 2020.

\bibitem[Sekuboyina et~al.(2021)Sekuboyina, Husseini, Bayat, L{\"o}ffler,
  Liebl, Li, Tetteh, Kuka{\v{c}}ka, Payer, {\v{S}}tern,
  et~al.]{SEKUBOYINA2021102166}
Anjany Sekuboyina, Malek~E Husseini, Amirhossein Bayat, Maximilian L{\"o}ffler,
  Hans Liebl, Hongwei Li, Giles Tetteh, Jan Kuka{\v{c}}ka, Christian Payer,
  Darko {\v{S}}tern, et~al.
\newblock Verse: a vertebrae labelling and segmentation benchmark for
  multi-detector ct images.
\newblock \emph{Medical image analysis}, 73:\penalty0 102166, 2021.

\bibitem[Mahadevan(2018)]{MAHADEVAN2018327}
Vishy Mahadevan.
\newblock Anatomy of the vertebral column.
\newblock \emph{Surgery (Oxford)}, 36\penalty0 (7):\penalty0 327--332, 2018.

\bibitem[Hines(2018)]{SpineAna11:online}
Tonya Hines.
\newblock Spine anatomy | mayfield brain \& spine.
\newblock \url{https://mayfieldclinic.com/pe-anatspine.htm}, 09 2018.

\bibitem[Bibb et~al.(2014)Bibb, Eggbeer, and Paterson]{bibb2014medical}
Richard Bibb, Dominic Eggbeer, and Abby Paterson.
\newblock \emph{Medical modelling: the application of advanced design and rapid
  prototyping techniques in medicine}.
\newblock Woodhead Publishing, 2014.

\bibitem[L{\"o}ffler et~al.(2020{\natexlab{a}})L{\"o}ffler, Sollmann, Mei,
  Valentinitsch, No{\"e}l, Kirschke, and Baum]{loffler2020x}
MT~L{\"o}ffler, N~Sollmann, K~Mei, A~Valentinitsch, PB~No{\"e}l, JS~Kirschke,
  and T~Baum.
\newblock X-ray-based quantitative osteoporosis imaging at the spine.
\newblock \emph{Osteoporosis Intern.}, 31\penalty0 (2):\penalty0 233--250,
  2020{\natexlab{a}}.

\bibitem[Zou et~al.(2019)Zou, Li, Deng, Du, and Xu]{zou2019use}
Da~Zou, Weishi Li, Chao Deng, Guohong Du, and Nanfang Xu.
\newblock The use of ct hounsfield unit values to identify the undiagnosed
  spinal osteoporosis in patients with lumbar degenerative diseases.
\newblock \emph{European Spine Journal}, 28\penalty0 (8):\penalty0 1758--1766,
  2019.

\bibitem[Kim et~al.(2019)Kim, Kim, Lee, Choi, Han, and Nam]{kim2019hounsfield}
Kyung~Joon Kim, Dong~Hwan Kim, Jae~Il Lee, Byung~Kwan Choi, In~Ho Han, and
  Kyoung~Hyup Nam.
\newblock Hounsfield units on lumbar computed tomography for predicting
  regional bone mineral density.
\newblock \emph{Open Medicine}, 14\penalty0 (1):\penalty0 545--551, 2019.

\bibitem[Lee et~al.(2013)Lee, Chung, Oh, and Park]{lee2013correlation}
Sungjoon Lee, Chun~Kee Chung, So~Hee Oh, and Sung~Bae Park.
\newblock Correlation between bone mineral density measured by dual-energy
  x-ray absorptiometry and hounsfield units measured by diagnostic ct in lumbar
  spine.
\newblock \emph{Journal of Korean Neurosurgical Society}, 54\penalty0
  (5):\penalty0 384, 2013.

\bibitem[Barron(2007)]{barron2007generation}
Valerie Barron.
\newblock Generation of a finite element model of the thoracolumbar spine.
\newblock \emph{Acta of bioengineering and biomechanics}, 9\penalty0
  (1):\penalty0 207, 2007.

\bibitem[Aroeira et~al.(2017)Aroeira, Pertence, Kemmoku, and
  Greco]{aroeira2017three}
Rozilene Maria~Cota Aroeira, Ant{\^o}nio Eust{\'a}quio de~Melo Pertence,
  Daniel~Takanori Kemmoku, and Marcelo Greco.
\newblock Three-dimensional geometric model of the middle segment of the
  thoracic spine based on graphical images for finite element analysis.
\newblock \emph{Research on Biomedical Engineering}, 33:\penalty0 97--104,
  2017.

\bibitem[Salsabili et~al.(2019)Salsabili, L{\'o}pez, and
  Barrio]{salsabili2019simplifying}
Neda Salsabili, Joaqu{\'\i}n~Santiago L{\'o}pez, and Mar{\'\i}a Isabel~Prieto
  Barrio.
\newblock Simplifying the human lumbar spine (l3/l4) material in order to
  create an elemental structure for the future modeling.
\newblock \emph{Australasian physical \& engineering sciences in medicine},
  42\penalty0 (3):\penalty0 689--700, 2019.

\bibitem[Campbell and Petrella(2016)]{campbell2016automated}
JQ~Campbell and AJ~Petrella.
\newblock Automated finite element modeling of the lumbar spine: using a
  statistical shape model to generate a virtual population of models.
\newblock \emph{Journal of biomechanics}, 49\penalty0 (13):\penalty0
  2593--2599, 2016.

\bibitem[Anitha et~al.(2020)Anitha, Baum, Kirschke, and
  Subburaj]{anitha2020effect}
D~Praveen Anitha, Thomas Baum, Jan~S Kirschke, and Karupppasamy Subburaj.
\newblock Effect of the intervertebral disc on vertebral bone strength
  prediction: A finite-element study.
\newblock \emph{The Spine Journal}, 20\penalty0 (4):\penalty0 665--671, 2020.

\bibitem[Castro-Mateos et~al.(2015)Castro-Mateos, Pozo, Perea{\~n}ez, Lekadir,
  Lazary, and Frangi]{castro2015statistical}
Isaac Castro-Mateos, Jose~M Pozo, Marco Perea{\~n}ez, Karim Lekadir, Aron
  Lazary, and Alejandro~F Frangi.
\newblock Statistical interspace models (sims): application to robust {3D}
  spine segmentation.
\newblock \emph{IEEE TTrans. on Medical Imaging}, 34\penalty0 (8):\penalty0
  1663--1675, 2015.

\bibitem[{\v{S}}tern et~al.(2011){\v{S}}tern, Likar, Pernu{\v{s}}, and
  Vrtovec]{vstern2011parametric}
Darko {\v{S}}tern, Bo{\v{s}}tjan Likar, Franjo Pernu{\v{s}}, and Toma{\v{z}}
  Vrtovec.
\newblock Parametric modelling and segmentation of vertebral bodies in {3D} ct
  and mr spine images.
\newblock \emph{Physics in Medicine \& Biology}, 56\penalty0 (23):\penalty0
  7505, 2011.

\bibitem[Laouissat et~al.(2018)Laouissat, Sebaaly, Gehrchen, and
  Roussouly]{laouissat2018classification}
F{\'e}thi Laouissat, Amer Sebaaly, Martin Gehrchen, and Pierre Roussouly.
\newblock Classification of normal sagittal spine alignment: refounding the
  roussouly classification.
\newblock \emph{European Spine Journal}, 27\penalty0 (8):\penalty0 2002--2011,
  2018.

\bibitem[Yeh et~al.(2021)Yeh, Weng, Huang, Fu, Tsai, and Yeh]{yeh2021deep}
Yu-Cheng Yeh, Chi-Hung Weng, Yu-Jui Huang, Chen-Ju Fu, Tsung-Ting Tsai, and
  Chao-Yuan Yeh.
\newblock Deep learning approach for automatic landmark detection and alignment
  analysis in whole-spine lateral radiographs.
\newblock \emph{Scientific reports}, 11\penalty0 (1):\penalty0 1--15, 2021.

\bibitem[Roussouly and Pinheiro-Franco(2011)]{roussouly2011sagittal}
Pierre Roussouly and Jo{\~a}o~Luiz Pinheiro-Franco.
\newblock Sagittal parameters of the spine: biomechanical approach.
\newblock \emph{European Spine Journal}, 20\penalty0 (5):\penalty0 578--585,
  2011.

\bibitem[Keller et~al.(2005)Keller, Colloca, Harrison, Harrison, and
  Janik]{keller2005influence}
Tony~S Keller, Christopher~J Colloca, Deed~E Harrison, Donald~D Harrison, and
  Tadeusz~J Janik.
\newblock Influence of spine morphology on intervertebral disc loads and
  stresses in asymptomatic adults: implications for the ideal spine.
\newblock \emph{The Spine Journal}, 5\penalty0 (3):\penalty0 297--309, 2005.

\bibitem[Lois~Zlolniski et~al.(2019)Lois~Zlolniski, Torres-Tamayo,
  Garc{\'\i}a-Mart{\'\i}nez, Blanco-P{\'e}rez, Mata-Escolano, Barash, Nalla,
  Martelli, Sanchis-Gimeno, and Bastir]{lois20193d}
Stephanie Lois~Zlolniski, Nicole Torres-Tamayo, Daniel
  Garc{\'\i}a-Mart{\'\i}nez, Esther Blanco-P{\'e}rez, Federico Mata-Escolano,
  Alon Barash, Shahed Nalla, Sandra Martelli, Juan~A Sanchis-Gimeno, and Markus
  Bastir.
\newblock {3D} geometric morphometric analysis of variation in the human lumbar
  spine.
\newblock \emph{American journal of physical anthropology}, 170\penalty0
  (3):\penalty0 361--372, 2019.

\bibitem[Casciaro and Massoptier(2007)]{casciaro2007automatic}
Sergio Casciaro and Laurent Massoptier.
\newblock Automatic vertebral morphometry assessment.
\newblock In \emph{2007 29th Annual Intern. Conference of the IEEE Engineering
  in Medicine and Biology Society}, pages 5571--5574. IEEE, 2007.

\bibitem[Shaw et~al.(2015)Shaw, Shaw, Cooperman, Eubanks, Li, and
  Kim]{shaw2015characterization}
Jeremy~D Shaw, Daniel~L Shaw, Daniel~R Cooperman, Jason~D Eubanks, Ling Li, and
  David~H Kim.
\newblock Characterization of lumbar spinous process morphology: a cadaveric
  study of 2,955 human lumbar vertebrae.
\newblock \emph{The Spine Journal}, 15\penalty0 (7):\penalty0 1645--1652, 2015.

\bibitem[Labelle et~al.(2011)Labelle, Aubin, Jackson, Lenke, Newton, and
  Parent]{labelle2011seeing}
Hubert Labelle, Carl-Eric Aubin, Roger Jackson, Larry Lenke, Peter Newton, and
  Stefan Parent.
\newblock Seeing the spine in {3D}: how will it change what we do?
\newblock \emph{Journal of Pediatric Orthopaedics}, 31:\penalty0 S37--S45,
  2011.

\bibitem[Fazzalari et~al.(2001)Fazzalari, Manthey, and
  Parkinson]{fazzalari2001intervertebral}
Nicola~L Fazzalari, Beverley Manthey, and Ian~H Parkinson.
\newblock Intervertebral disc disorganisation and its relationship to age
  adjusted vertebral body morphometry and vertebral bone architecture.
\newblock \emph{The Anatomical Record: An Official Publication of the American
  Association of Anatomists}, 262\penalty0 (3):\penalty0 331--339, 2001.

\bibitem[Edelsbrunner and Harer(2010)]{edelsbrunner2010computational}
Herbert Edelsbrunner and John Harer.
\newblock \emph{Computational topology: an introduction}.
\newblock American Mathematical Soc., 2010.

\bibitem[Edelsbrunner and M{\"u}cke(1994)]{edelsbrunner1994three}
Herbert Edelsbrunner and Ernst~P M{\"u}cke.
\newblock Three-dimensional alpha shapes.
\newblock \emph{ACM Trans. on Graphics (TOG)}, 13\penalty0 (1):\penalty0
  43--72, 1994.

\bibitem[Zhou and Yan(2010)]{zhou2010discriminatory}
Weiqiang Zhou and Hong Yan.
\newblock A discriminatory function for prediction of protein--dna interactions
  based on alpha shape modeling.
\newblock \emph{Bioinformatics}, 26\penalty0 (20):\penalty0 2541--2548, 2010.

\bibitem[DenOtter and Schubert(2022)]{Hounsfie99:online}
Tami~D. DenOtter and Johanna. Schubert.
\newblock Hounsfield unit - statpearls - ncbi bookshelf.
\newblock \url{https://www.ncbi.nlm.nih.gov/books/NBK547721/}, January 2022.

\bibitem[L{\"o}ffler et~al.(2020{\natexlab{b}})L{\"o}ffler, Sekuboyina, Jacob,
  Grau, Scharr, El~Husseini, Kallweit, Zimmer, Baum, and
  Kirschke]{doi:10.1148/ryai.2020190138}
Maximilian~T. L{\"o}ffler, Anjany Sekuboyina, Alina Jacob, Anna-Lena Grau,
  Andreas Scharr, Malek El~Husseini, Mareike Kallweit, Claus Zimmer, Thomas
  Baum, and Jan~S. Kirschke.
\newblock A vertebral segmentation dataset with fracture grading.
\newblock \emph{Radiology: Artificial Intelligence}, 2\penalty0 (4):\penalty0
  e190138, 2020{\natexlab{b}}.

\bibitem[Liebl et~al.(2021)Liebl, Schinz, Sekuboyina, Malagutti, L{\"o}ffler,
  Bayat, Husseini, Tetteh, Grau, Niederreiter, et~al.]{liebl2021computed}
Hans Liebl, David Schinz, Anjany Sekuboyina, Luca Malagutti, Maximilian~T
  L{\"o}ffler, Amirhossein Bayat, Malek~El Husseini, Giles Tetteh, Katharina
  Grau, Eva Niederreiter, et~al.
\newblock A computed tomography vertebral segmentation dataset with anatomical
  variations and multi-vendor scanner data.
\newblock \emph{arXiv preprint arXiv:2103.06360}, 2021.

\end{thebibliography}





% \end{thebibliography}

\end{document}